\documentclass[10pt,aps,twocolumn,prd,noshowpacs,nofootinbib,noshowkeys,floatfix,superscriptaddress]{revtex4}
\usepackage[dvips]{graphics,graphicx}
\usepackage[colorlinks=true,linktocpage=true,linkcolor=blue,citecolor=blue]{hyperref}
\usepackage[usenames,dvipsnames]{color}
\usepackage{amsmath, amssymb,oldgerm}
\usepackage{multirow}
\usepackage{longtable}
\usepackage{color}
\usepackage[normalem]{ulem}  
\usepackage{braket}
\usepackage[mathscr]{eucal}
\newcommand{\beq}{\begin{equation}}
\newcommand{\eeq}{\end{equation}}

\newcommand{\Tr}{\text{Tr}}
\newcommand{\n}{\nonumber}
\newcommand{\x}{x_1}
\newcommand{\y}{x_2}
\newcommand{\s}{y}
\newcommand{\X}{x}
\newcommand{\F}{\mathcal{F}}
\newcommand{\Pc}{\mathcal{P}}
\newcommand{\V}{\mathcal{V}}
\newcommand{\A}{\mathcal{A}}
\newcommand{\Sc}{\mathcal{S}}

\begin{document}

\preprint{}

\title{Kinetic theory \textcolor{black}{for} massive spin-1/2 particles from the Wigner-function formalism}

\author{Nora Weickgenannt}

\affiliation{Institute  for  Theoretical  Physics,  Goethe  University,
Max-von-Laue-Str.\  1,  D-60438  Frankfurt  am  Main,  Germany}

\author{Xin-li Sheng}

\affiliation{Institute  for  Theoretical  Physics,  Goethe  University,
Max-von-Laue-Str.\  1,  D-60438  Frankfurt  am  Main,  Germany}
\affiliation{Interdisciplinary Center for Theoretical Study and Department of
Modern Physics, University of Science and Technology of China, Hefei,
Anhui 230026, China}

\author{Enrico Speranza}

\affiliation{Institute  for  Theoretical  Physics,  Goethe  University,
Max-von-Laue-Str.\  1,  D-60438  Frankfurt  am  Main,  Germany}

\author{Qun Wang}

\affiliation{Interdisciplinary Center for Theoretical Study and Department of
Modern Physics, University of Science and Technology of China, Hefei,
Anhui 230026, China}

\author{Dirk H.\ Rischke}

\affiliation{Institute  for  Theoretical  Physics,  Goethe  University,
Max-von-Laue-Str.\  1,  D-60438  Frankfurt  am  Main,  Germany}
\affiliation{Interdisciplinary Center for Theoretical Study and Department of
Modern Physics, University of Science and Technology of China, Hefei,
Anhui 230026, China}

\begin{abstract}
We calculate the Wigner function for
massive spin-1/2 particles in an inhomogeneous electromagnetic field to leading order in the Planck constant $\hbar$. 
Going beyond leading order in $\hbar$ we then derive a generalized Boltzmann equation 
in which the force 
exerted by an inhomogeneous electromagnetic field on the particle dipole moment 
arises naturally. Furthermore, a kinetic equation for this dipole moment is derived. Carefully taking the massless
limit we find agreement with previous results.
The case of global equilibrium with rotation is also studied. 
Finally, we outline the derivation of fluid-dynamical equations from the components of the Wigner function. 
The conservation of total angular momentum is promoted as an additional fluid-dynamical
equation of motion.
Our framework can be used to study polarization effects induced by vorticity and magnetic field in 
relativistic heavy-ion collisions.
\end{abstract}



\maketitle

\section{Introduction}

\textcolor{black}{Relativistic heavy-ion collisions (HICs) create a new phase of hot and dense strong-interaction matter,
the quark-gluon plasma (QGP) [see e.g.\ Ref.\ \cite{Proceedings:2019drx}]. 
The interaction rates between its constituents are sufficiently large
that the matter rapidly reaches a state which can be described by fluid dynamics
\cite{Kurkela:2018oqw}.
In non-central HICs the global angular momentum generates
a non-vanishing vorticity of the QGP fluid. Furthermore, 
in such collisions} a strong magnetic field is formed due to the
electric current produced by the spectator protons constituting the colliding ions. 

\textcolor{black}{In the QGP,
quarks can be considered as (nearly) massless fermions. The} 
interplay between the chiral anomaly on the one hand and the
magnetic field and the fluid \textcolor{black}{vorticity} 
on the other hand gives rise to novel transport phenomena called chiral effects.
Two such phenomena are the chiral magnetic effect (CME) 
\cite{Kharzeev:2007jp} and \textcolor{black}{the} chiral vortical effect (CVE) \cite{Son:2009tf},
where a \textcolor{black}{charge} current is induced along the \textcolor{black}{direction of the magnetic field and 
the vorticity}, 
respectively. 
\textcolor{black}{Large-scale experimental efforts are currently under way
to discover these phenomena in HICs  [for a recent review, see Ref.~\cite{Kharzeev:2015znc}].}

\textcolor{black}{From the theoretical point of view, it is therefore mandatory to develop a theory which 
allows to study such transport phenomena in chiral fluids.
One approach is chiral kinetic theory, which has been derived using various methods, e.g.\ the}
classical action~\cite{Son:2012wh,Stephanov:2012ki,Son:2012zy,Chen:2012ca,Manuel:2013zaa,Chen:2014cla,Manuel:2014dza,Chen:2015gta,Gorbar:2017cwv}, the
Wigner function~\cite{Hidaka:2016yjf,Hidaka:2017auj,Huang:2018wdl,Gao:2018wmr,Yang:2018lew,Gao:2018jsi}, 
and the world-line
formalism~\cite{Mueller:2017lzw,Mueller:2017arw,Mueller:2019gjj}. 
\textcolor{black}{In Refs.~\cite{Hidaka:2016yjf,Hidaka:2017auj} it was shown} that, 
using Wigner functions, one is able to recover
the ``side-jump" phenomenon first \textcolor{black}{discussed} 
in Refs.\ \cite{Chen:2014cla,Chen:2015gta} in order to ensure
total angular-momentum conservation in binary collisions. Furthermore, the inclusion of the chiral effects in fluid
dynamics was studied in Refs.~\cite{Son:2009tf,Neiman:2010zi,Sadofyev:2010pr}.

Another intriguing phenomenon occurring in the rotating QGP is that particles in the medium can \textcolor{black}{be}
polarized in a way resembling the Einstein-de Haas~\cite{dehaas:1915} and Barnett effects~\cite{Barnett:1935}. Recently,
the STAR Collaboration presented \textcolor{black}{experimental evidence for the alignment of}
the spin of $\Lambda$ hyperons
\textcolor{black}{with} the global angular momentum in peripheral \textcolor{black}{HICs}~\cite{STAR:2017ckg}. 
This finding revealed, for the first time,
the strong vortical structure of the QGP. Many theoretical works have explored spin-polarization mechanisms triggered by
vorticity in HICs. In particular, the importance of the spin-orbit interaction~\cite{Liang:2004ph,Gao:2007bc,Chen:2008wh}
and the relation between spin polarization and thermal vorticity
in \textcolor{black}{local thermodynamical} equilibrium have been
studied~\cite{Becattini:2007sr,Becattini:2013fla,Becattini:2013vja,Becattini:2016gvu}. A fluid-dynamical
description, which includes the space-time evolution of the spin polarization, was proposed in
Refs.~\cite{Florkowski:2017ruc,Florkowski:2017dyn,Florkowski:2018myy,Florkowski:2018fap}. 
However, this formulation is based on
a specific choice \textcolor{black}{for} the energy-momentum and spin tensors. The physical implications of different sets of
energy-momentum and spin tensors in fluid dynamics was investigated in Ref.~\cite{Becattini:2018duy}.

Although there has been intense theoretical activity which has led to a deeper understanding of 
\textcolor{black}{the} transport properties of
chiral matter, few studies have \textcolor{black}{attempted to derive} 
a covariant kinetic \textcolor{black}{theory} for \textit{massive} particles
using Wigner functions~\cite{Fang:2016vpj,Florkowski:2018ahw}. The aim of this paper is to 
\textcolor{black}{fill this gap. We} derive kinetic theory
for massive spin-1/2 particles in an inhomogeneous electromagnetic field as a basis
to study polarization effects in HICs. Our starting point is the covariant formulation of the Wigner
function~\cite{Heinz:1983nx,Elze:1986qd,Vasak:1987um,Zhuang:1995pd,
Florkowski:1995ei,Blaizot:2001nr,Wang:2001dm}. In order to solve the equations of motion for the Wigner function,
we employ an expansion in the Planck constant $\hbar$ and truncate at the lowest non-trivial order.
This approximation is valid if the following two assumptions hold: 
\begin{itemize}
\item[(i)] $\hbar | \gamma^\mu\nabla_\mu 
W| \ll m|W|$,
where $W$ is the Wigner function, \textcolor{black}{$m$ is the particle mass,} 
$\nabla_\mu$ represents the gradient operator in Eq.\ (\ref{nabla})~\cite{Vasak:1987um,DeGroot:1980dk}, and  
the modulus applies to each component of the corresponding matrix in Dirac space, 
\item[(ii)] $\hbar \ll \Delta R \Delta P$,
where $\Delta R$ is a spatial scale over which the electromagnetic field tensor varies significantly and $\Delta P$
a momentum scale over which the Wigner function varies significantly.
\end{itemize}
Assumption (i) implies that the \textcolor{black}{$\hbar$--expansion} is effectively a \textcolor{black}{gradient} expansion.
Assumption (ii) \textcolor{black}{allows to truncate the}
power-series expansion of the Bessel functions entering the equations of motion of the
Wigner function~\cite{Vasak:1987um}.

Under these assumptions, we first give an explicit derivation of the leading-order solution.
Then, considering the equation of motion for the Wigner function to first and second order in 
$\hbar$, we derive a generalized Boltzmann equation, where the external force acting on the particles is given by
two contributions. The first one is the Lorentz force, which gives rise to the usual Vlasov term, and the second one is the
Mathisson force~\cite{Bailey:1975fe}, i.e., the force exerted on \textcolor{black}{the particle's dipole moment 
in an inhomogeneous}
electromagnetic field. \textcolor{black}{In our context, the dipole moment arises} from the spin of the particle.
We show how to \textcolor{black}{take the massless limit, obtaining 
a result that} agrees with previous works~\cite{Hidaka:2016yjf,Hidaka:2017auj}.
We also study the solution of the Boltzmann equation in the case of global equilibrium with rigid rotation.
Finally, \textcolor{black}{we derive} fluid-dynamical equations of motion with spin degrees of freedom from
the Wigner function using the canonical definitions of \textcolor{black}{the} 
energy-momentum and spin tensors. In accordance with previous
works \cite{Becattini:2018duy,Florkowski:2017ruc}, the conservation of the total angular momentum is promoted as
an additional fluid-dynamical equation, where the divergence of the spin tensor is related to the antisymmetric part
of the energy-momentum tensor.

\textcolor{black}{We use units $c=k_B=1$ throughout this 
paper}. It is useful to explicitly keep \textcolor{black}{Planck's} constant $\hbar$, \textcolor{black}{since it will be} our
power-counting parameter. The convention for the metric tensor is $g^{\mu \nu} = \mathrm{diag} (+1,-1,-1,-1)$ 
\textcolor{black}{and $\epsilon^{0123}=-\epsilon_{0123}=+1$ for the rank-four Levi-Civita tensor}. We
use the notation $a^\mu b_\mu \equiv a \cdot b$ for the scalar product of two four-vectors $a^\mu, b^\mu$ and
$\mathbf{a} \cdot \mathbf{b}$ for the corresponding scalar product of two spatial vectors $\mathbf{a}, \mathbf{b}$.
\textcolor{black}{A two-dimensional vector in spin space is} denoted by $\vec{a}$.
The electromagnetic four-potential is $\mathbb{A}^\mu$, where the electromagnetic charge is
absorbed into its definition.
We denote the dipole-moment tensor as $\Sigma^{\mu\nu}$. This quantity corresponds to
the spin tensor $S^{\mu \nu}$ of Refs.~\cite{Chen:2014cla,Chen:2015gta}.
In this paper the term ``spin tensor'' is reserved for the \textcolor{black}{rank-three} Lorentz tensor~$S^{\lambda, \mu\nu}$.

\section{Equations for the Wigner function for massive fermions}

The Wigner function is defined as the Fourier transform of the two-point correlation function \cite{Elze:1986qd},
\begin{eqnarray}
W_{\alpha\beta}(\X,p)& =& \int \frac{d^4 \s}{(2\pi\hbar)^4} e^{-\frac{i}{\hbar}p\cdot \s}  \n \\
& &  \times \left\langle :\bar{\psi}_\beta(\x)U(\x,\y)\psi_\alpha (\y): \right\rangle\,. \label{Wignerdef}
\end{eqnarray}
Here, $\x$ and $\y$ are the space-time coordinates of two different points, with $\s^\mu\equiv \x^\mu-\y^\mu$ and
$\X^\mu\equiv (\x^\mu+\y^\mu)/2$.
The gauge link is defined as
\begin{equation}
 U(\x,\y)=\exp \left[-\frac{i}{\hbar}\s^\mu\int_{-1/2}^{1/2} dt\, \mathbb{A}_\mu(x+t\s)\right]\,. \label{gaugelink}
\end{equation}
In this paper, $\mathbb{A}_\mu$ will be treated as an external, classical field (otherwise, the gauge link would
need to be path-ordered). The particular choice of path for the integration between $\x$ and $\y$ ensures that
$p^\mu$ is the kinetic momentum. Note that the factors $2 \pi \hbar$ in the denominator in Eq.\ (\ref{Wignerdef})
belong to the phase-space volume and do not participate in the $\hbar$--counting employed throughout this paper.

Starting from the Dirac equation and its adjoint,
\begin{equation} \label{DiracEq}
 (i\hbar\gamma \cdot D-m)\psi=0 = \bar{\psi}(i\hbar\gamma \cdot D^\dagger+m)\,,
\end{equation}
where $D_\mu \equiv \partial_{x\mu} +\frac{i}{\hbar} \mathbb{A}_\mu$ is the covariant derivative,
one can derive the kinetic equation for the Wigner function as \cite{Elze:1986qd}
\begin{equation}
(\gamma \cdot K - m)W(x,p)=0\,. \label{Dirac}
\end{equation}
Here one has defined the operator
\begin{equation}
K^\mu\equiv \Pi^\mu+\frac12i\hbar\nabla^\mu\,,
\end{equation}
with the generalized space-time derivative and momentum operators
\begin{eqnarray}
\nabla^\mu&\equiv&\partial_x^\mu-j_0(\Delta)F^{\mu\nu}\partial_{p\nu}\,, \label{nabla}\\
\Pi^\mu&\equiv& p^\mu-\frac{\hbar}{2} j_1(\Delta)F^{\mu\nu}\partial_{p\nu}\,, \label{pW}
\end{eqnarray}
where $\Delta \equiv \frac{\hbar}{2}\, \partial_p\cdot\partial_x$ and $F^{\mu \nu} = \partial^\mu_x \mathbb{A}^\nu
- \partial^\nu_x \mathbb{A}^\mu$ is the electromagnetic
field-strength tensor. We should emphasize that in Eq.\ (\ref{Dirac}) the space-time derivative
$\partial_x$ contained in $\Delta$ only acts on  $F^{\mu\nu}$, but not on the Wigner function.
The functions $j_0(x)=\sin x /x$ and $j_1(x)=(\sin x-x \cos x) /x^2$ are spherical Bessel functions.
If we assume that the particles only interact with the classical electromagnetic field but not among themselves
(which, in the language of kinetic theory, is the limit of the collisionless Boltzmann-Vlasov equation),
Eq.\ (\ref{Dirac}) is exact and contains the full dynamics of the Wigner function.

In order to derive a kinetic equation for massive spin-1/2 particles, it is advantageous to decompose the Wigner function
in terms of a basis formed by the 16 independent generators of the Clifford algebra $ \{1, \gamma^5, \gamma^\mu,
\gamma^5\gamma^\mu, \sigma^{\mu\nu}\}$, with $\gamma^5\equiv i\gamma^0\gamma^1\gamma^2\gamma^3$
and $\sigma^{\mu\nu}\equiv\frac i2[\gamma^\mu,\gamma^\nu]$,
\begin{equation}
W=\frac14\left(\F+i\gamma^5\Pc+\gamma^\mu \V_\mu+\gamma^5\gamma^\mu \A_\mu
+\frac12\sigma^{\mu\nu}\Sc_{\mu\nu}\right)\,. \label{dec}
\end{equation}
The coefficients $\F, \Pc, \V^\mu, \A^\mu,$ and $\Sc^{\mu\nu}$ are real functions of the phase-space coordinates ${x,p}$
and correspond to the scalar, pseudo-scalar, vector, axial-vector, and tensor components of the Wigner function.
Some of them have an obvious physical meaning \cite{BialynickiBirula:1991tx}. For example, $\V^\mu$ is the
fermion four-current and $\A^\mu$ is related to the spin density. Using the trace properties of the Dirac matrices,
the coefficients in Eq.\ (\ref{dec}) are given by
\begin{eqnarray}
\F&=&\Tr(W)\,, \n \\
\Pc&=&-i\,\Tr(\gamma^5W)\,,\n\\
\V^\mu&=&\Tr(\gamma^\mu W)\,,\n\\
\A^\mu&=&\Tr(\gamma^\mu\gamma^5W)\,,\n\\
\Sc^{\mu\nu}&=&\Tr(\sigma^{\mu\nu}W)\,. \label{coeff}
\end{eqnarray}
Replacing $W$ in Eq.\ (\ref{Dirac}) by the decomposition (\ref{dec}), we find the following complex-valued equations:
\begin{eqnarray}
K \cdot \V -m\F&=&0\,,\n \\
K \cdot \A+im\Pc&=&0\,,\n \\
K_\mu \F+iK^\nu \Sc_{\nu\mu}-m\V_\mu&=&0\,, \n\\
iK_\mu \Pc+\frac12 \epsilon_{\mu\nu\alpha\beta}K^\nu \Sc^{\alpha\beta}+m \A_\mu&=&0\,,\n\\
-iK_{[\mu} \V_{\nu]}-\epsilon_{\mu\nu\alpha\beta}K^\alpha \A^\beta-m\Sc_{\mu\nu}&=&0\,,
\end{eqnarray}
where $A_{[\mu}B_{\nu]}\equiv A_\mu B_\nu-A_\nu B_\mu$. Decomposing these equations into
their real and imaginary parts, we obtain a set of coupled equations which determine the coefficients
in the decomposition (\ref{dec}) of the Wigner function. The real parts read
\begin{eqnarray}
\Pi\cdot \V -m\F&=&0\,,\label{F}\\
\frac{\hbar}{2}\nabla\cdot  \A+m\Pc&=&0\,,\label{P}\\
\Pi_\mu \F-\frac{\hbar}{2}\nabla^\nu \Sc_{\nu\mu}-m\V_\mu&=&0\,,\label{V}\\
-\frac{\hbar}{2}\nabla_\mu \Pc+\frac12\epsilon_{\mu\nu\alpha\beta}\Pi^\nu S^{\alpha\beta}+m\A_\mu&=&0\,,\label{A}\\
\frac{\hbar}{2}\nabla_{[\mu} \V_{\nu]}-\epsilon_{\mu\nu\alpha\beta}\Pi^\alpha \A^\beta-m\Sc_{\mu\nu}&=&0\,,\label{S}
\end{eqnarray}
and the imaginary parts are
\begin{eqnarray}
\hbar\nabla \cdot \V&=&0\,,\label{Vkin}\\
\Pi \cdot \A&=&0\,,\label{orth}\\
\frac{\hbar}{2}\nabla_\mu \F+\Pi^\nu \Sc_{\nu\mu}&=&0\,,\label{B}\\
\Pi_{\mu}\Pc+\frac{\hbar}{4}\epsilon_{\mu\nu\alpha\beta}\nabla^\nu \Sc^{\alpha\beta}&=&0\,,\label{Skin}\\
\Pi_{[\mu} \V_{\nu]}+\frac{\hbar}{2}\epsilon_{\mu\nu\alpha\beta}\nabla^\alpha \A^\beta&=&0\,. \label{Akin}
\end{eqnarray}
In the next sections, we will explicitly solve Eqs.\ (\ref{F}) -- (\ref{Akin}) to zeroth order in $\hbar$, and
then derive kinetic equations which the general solution has to fulfill up to first order in $\hbar$.

\section{Zeroth-order solution}
\label{zerothorderWigner_direct}

To zeroth order in $\hbar$, the operator $K^\mu = p^\mu$ and Eq.\ (\ref{Dirac}) reduces to
\begin{equation}
(\gamma \cdot p-m)W^{(0)}(\X,p)=0\,.
\end{equation}
The solution is given by \cite{DeGroot:1980dk,Fang:2016vpj}
\begin{equation} \label{solution_hbar=0}
 W_{\alpha\beta}^{(0)}(x,p)=W_{\alpha\beta}^+(x,p)+W_{\alpha\beta}^-(x,p)\,,
\end{equation}
\textcolor{black}{where}
\begin{eqnarray}
 W_{\alpha\beta}^+(x,p)&=&\frac{1}{(2\pi \hbar)^3}\int d^4q\, \theta(q^0)\delta(q^2-m^2)\delta^4(p-q)\n \\ &&
 \times \sum_{rs} u_\alpha(\mathbf{q},r)\bar{u}_\beta(\mathbf{q},s)f_{rs}^{(0)+}(x,\mathbf{q}) \,, \label{W+} \\
 W_{\alpha\beta}^-(x,p)&=&- \frac{1}{(2\pi \hbar)^3 }\int d^4q\, \theta(q^0)\delta(q^2-m^2)\delta^4(p+q)\n\\ &&
 \times \sum_{rs} v_\alpha(\mathbf{q},s)\bar{v}_\beta(\mathbf{q},r)f_{sr}^{(0)-} (x,\mathbf{q}) \label{W-}
\end{eqnarray}
\textcolor{black}{are the contributions} from positive and negative energies, respectively. Here,
$f_{rs}^{+}(x,\mathbf{q})$ and
$f_{sr}^{-}(x,\mathbf{q})$ are the distribution functions for
fermions and anti-fermions, respectively,
which are in general matrices in spin space.
The spin indices label spin states parallel, $r,s=+$, or anti-parallel, $r,s=-$, to the quantization direction $\mathbf{s}$
in the rest frame of the particle, respectively. 

This spin quantization direction can in principle be chosen arbitrarily. However,
the most convenient choice is to quantize the spin with respect to the polarization direction
\cite{Vasak:1987um,Fang:2016vpj}. 
In other words, we choose a spin basis in which the new distribution functions $\tilde{f}_{rs}^\pm$ are diagonal, i.e.,
\begin{equation}
 \tilde{f}_{rs}^{(0)\pm}=f^{(0)\pm}_{s}\delta_{rs} \,.\label{diagrs}
\end{equation}
\textcolor{black}{In App.\ \ref{XX} we demonstrate} that such a choice is always possible, \textcolor{black}{at the expense
of introducing space-time dependent spinors, cf.\ Eq.\ (\ref{tilde_u}).
We will} also use the diagonal basis in the calculation of \textcolor{black}{the contributions of higher order in $\hbar$} 
in the following sections. 

\textcolor{black}{As shown in App.\ \ref{XX},} the spin quantization direction
$n^{(0)\mu}$ is given by
\begin{equation} \label{nmu}
 n^{(0)\mu}(x,p)\equiv \theta(p^0)n^{+\mu}(x,\mathbf{p})-\theta(-p^0)n^{-\mu}(x,\mathbf{p})\,,
\end{equation}
\textcolor{black}{where
\begin{eqnarray}
 n^{+\mu}(x,\mathbf{p})=\left(\frac{\mathbf{n}^+\cdot\mathbf{p}}{m},\,
 \mathbf{n}^++\frac{\mathbf{n}^+\cdot \mathbf{p}}{m(m+E_\mathbf{p})}\,\mathbf{p}\right)\,,\n\\
 n^{-\mu}(x,\mathbf{p})=\left(\frac{\mathbf{n}^-\cdot\mathbf{p}}{m},\,
 -\mathbf{n}^--\frac{\mathbf{n}^-\cdot \mathbf{p}}{m(m+E_\mathbf{p})}\,\mathbf{p}\right)\! .
 \label{nDef}
\end{eqnarray}
Here, $\mathbf{n}^\pm$ is the spin quantization direction in the rest frame
of the particle/anti-particle [cf.\ Eq.\ (\ref{spinqu})] and $E_\mathbf{p}=\sqrt{\mathbf{p}^{\,2} + m^2}$.
The spin quantization direction $\mathbf{n}^\pm$ transforms as an axial vector
under Lorentz boosts and parity transformations.
We show in App.\ \ref{XX} that $\mathbf{n}^\pm$ depends in general on $\mathbf{p}$ and $x$, thus $n^{\pm\mu}$
is defined locally. The vector $n^{(0)\mu}$ is aligned with the polarization direction and 
agrees} with the classical spin vector, i.e., as we will see later, it
obeys the classical equation for spin precession in an electromagnetic field, the so-called
Bargmann--Michel--Telegdi (BMT) equation \cite{Bargmann:1959gz}.
Moreover, \textcolor{black}{$n^{(0)\mu}$ fulfills $p\cdot n^{(0)}=0$}
(which can be seen \textcolor{black}{using Eqs.\ (\ref{u_identity}) and (\ref{v_identity}) and applying}
the Dirac equation for the $u$-- and $v$--spinors \textcolor{black}{as well as} the identity
$\bar{u}(\mathbf{p},r) \gamma^5 u(\mathbf{p},s)=\bar{v}(-\mathbf{p},r) \gamma^5 v(-\mathbf{p},s)=0$).

Equations (\ref{solution_hbar=0}) -- (\ref{W-}) represent the solution obtained in Ref.\ \cite{DeGroot:1980dk} for vanishing
electromagnetic fields. However, this is also the solution for non-vanishing electromagnetic fields,
since the form of Eq.\ (\ref{Dirac}) remains the same. The momentum variable $p^\mu$ is then the kinetic (and not
the canonical) momentum.

Closer inspection of Eq.\ (\ref{Dirac}) reveals that Eq.\ (\ref{solution_hbar=0}) with Eqs.\ (\ref{W+}), (\ref{W-}) is 
\textcolor{black}{also
a solution to Eq.\ (\ref{Dirac}) at arbitrary order in $\hbar$,}
if $\gamma \cdot \nabla W^{(0)} =0$ and $\gamma_\mu F^{\mu \nu} \partial_{p \nu} W^{(0)} =0$ (because then the
$\hbar$--dependence of the operator $K^\mu$ vanishes).
In the absence of electromagnetic fields, one \textcolor{black}{at least} needs to require that 
$\gamma \cdot \partial_x W^{(0)} =0$. 
In the \textcolor{black}{full solution, i.e., the solution to all orders in $\hbar$,} the momentum variable
$q$ is no longer equal to the kinetic momentum $p$. This is
obviously not the case for Eqs.\ (\ref{W+}), (\ref{W-}), since they are proportional to $\sim \delta^4(p\mp q)$,
see also the discussion in Ref.\ \cite{DeGroot:1980dk}.

Now we easily obtain the coefficients of the decomposition (\ref{dec}) using Eqs.\ (\ref{coeff}) and (\ref{diagrs}). We find
\begin{eqnarray}
 \F^{(0)}(x,p)&=&m\,\delta(p^2-m^2)V^{(0)}(x,p)\,,\n\\
 \Pc^{(0)}(x,p)&=&0\,,\n\\
 \V^{(0)}_\mu(x,p)&=&p_\mu\delta(p^2-m^2)V^{(0)}(x,p)\,,\n\\
 \A^{(0)}_\mu(x,p)&=&m\,n_\mu^{(0)}(x,\mathbf{p}) \delta(p^2-m^2)A^{(0)}(x,p)\,,\n\\
 \Sc^{(0)}_{\mu\nu}(x,p)&=& m\,\Sigma^{(0)}_{\mu\nu}(x,\mathbf{p})\delta(p^2-m^2)A^{(0)}(x,p)\,, \label{Wignerdirect}
\end{eqnarray}
with
\begin{equation}
V^{(0)}(x,p) \equiv \frac{2}{(2\pi \hbar)^3}\sum_{es} \theta(e\,p^0)f^{(0)e}_s(x,e\,\mathbf{p}) \label{V_Wignerdirect}
\end{equation}
and
\begin{equation}
A^{(0)}(x,p) \equiv \frac{2}{(2\pi \hbar)^3}\sum_{es}s\, \theta(e\,p^0)f^{(0)e}_s(x,e\,\mathbf{p}) \label{A_Wignerdirect}
\end{equation}
 where $e=\pm$, $f^{(0)e}_s$ are the distribution functions in the diagonal basis, and the dipole-moment tensor 
\textcolor{black}{is defined as}
\begin{equation}
\Sigma_{\mu\nu}^{(0)}(x,p) =-\frac{1}{m}\epsilon_{\mu\nu\alpha\beta}p^\alpha n^{(0)\beta}\,, \label{Sigmadirect}
\end{equation}
for the proof, see App.\ \ref{XX}.

\section{General solution up to order $\hbar$}

In this section we derive the general solution for Eqs.\ (\ref{F}) -- (\ref{Akin}) \textcolor{black}{to first order
in $\hbar$}. We emphasize that these
equations are not independent from each other. We prove in App.\ \ref{prove} that Eq.\ (\ref{B}) can be derived from
Eqs.\ (\ref{F}), (\ref{S}), (\ref{Vkin}), (\ref{Akin}), and Eq.\ (\ref{Skin}) can be derived from Eqs.\ (\ref{P}), (\ref{S}),
(\ref{orth}), (\ref{Akin}). Thus, one can ignore Eqs.\ (\ref{B}), (\ref{Skin}) when solving this system of
partial differential equations.

Using Eqs.\ (\ref{P}), (\ref{V}), (\ref{A}) one can express the pseudo-scalar, vector, and axial-vector parts $\Pc$,
$\V^\mu$, and $\A^{\mu}$ as follows:
\begin{eqnarray}
\Pc&=&-\frac{\hbar}{2m}\nabla \cdot \A \,, \n\\
\V_\mu&=& \frac{1}{m}\Pi_\mu \F-\frac{\hbar}{2m}\nabla^\nu \Sc_{\nu\mu}\,, \n\\
\A_{\mu}&=&\frac{\hbar}{2m}\nabla_{\mu}\Pc -\frac{1}{2m}\epsilon_{\mu\nu\alpha\beta}\Pi^\nu \Sc^{\alpha \beta }\,.
\label{PVS}
\end{eqnarray}
Inserting them back into Eqs.\ (\ref{F}), (\ref{S}) one obtains the modified on-shell conditions for the scalar and
tensor components,
\begin{eqnarray}
(\Pi \cdot \Pi-m^2)\F&=&\frac{\hbar}{2}\Pi^\mu\nabla^\nu \Sc_{\nu\mu}\,, \n\\
(\Pi \cdot \Pi-m^2)\Sc_{\mu\nu}&=&-\Pi^\alpha \Pi_{[\mu}\Sc_{\nu]\alpha}
-\frac{\hbar}{2}\nabla_{[\mu}\Pi_{\nu]}\F \n \\ && -\frac{\hbar^2}{4}\nabla_{[\mu} \nabla^\alpha  \Sc_{\nu]\alpha}
+ \frac{\hbar}{2}\epsilon_{\mu\nu\alpha\beta}\Pi^\alpha \nabla^\beta\Pc\,. \n \\ \label{on-shell}
\end{eqnarray}
Equations (\ref{PVS}), (\ref{on-shell}) are equivalent to Eqs.\ (\ref{F}) -- (\ref{S}). In general, the right-hand sides
are non-vanishing, which indicates that the Wigner function contains off-shell effects.

From their definitions (\ref{nabla}), (\ref{pW}), we observe that the operators $\nabla^{\mu}$ and $\Pi^{\mu}$
can be expanded in a series of powers in $\hbar^2$. In order to derive the semi-classical limit, we
may truncate these series at order $\hbar^0$ and $\hbar^2$, respectively,
\begin{eqnarray}
\nabla^\mu&=& \sum_{n=0}^\infty \hbar^{2n}\nabla^{(2n)\mu}
= \nabla^{(0)\mu}+\mathcal{O}(\hbar^2) \,, \n \\
\Pi^\mu&=&\sum_{n=0}^\infty \hbar^{2n}\Pi^{(2n)\mu} \n \\
& = & p^\mu-\frac{\hbar^2}{12}(\partial_x^\alpha F^{\mu\nu})
\partial_{p\nu} \partial_{p\alpha}+\mathcal{O}(\hbar^4) \,,
\end{eqnarray}
where $\nabla^{(0)\mu}\equiv \partial_x^\mu -F^{\mu\nu}\partial_{p\nu}$.
We also expand the functions $\F, \Pc, \V^\mu, \A^\mu, \Sc^{\mu\nu}$ into power series in $\hbar$, e.g.,\
\begin{equation}
\F=\sum_{n=0}^\infty \hbar^n \F^{(n)}\, .
\end{equation}
Inserting these expansions into Eqs.\ (\ref{F}) -- (\ref{Akin}) and then comparing order by order in $\hbar$
one can get a set of equations which we will analyze up to second order in $\hbar$ in the remainder of this section.

\subsection{Zeroth order in $\hbar$}

We first analyze the on-shell conditions (\ref{on-shell}) for the scalar and tensor components
to leading order in $\hbar$ and show that the direct calculation of the Wigner function to this order
presented in Sec.\ \ref{zerothorderWigner_direct}
is consistent with these conditions. To order $\mathcal{O}(\hbar^0)$, Eq.\ (\ref{on-shell}) reads
\begin{eqnarray}
(p^2-m^2)\F^{(0)}&=&0 \, , \n\\
(p^2-m^2)\Sc^{(0)}_{\mu\nu} &=& 0\, ,
\end{eqnarray}
where we have used $p^\nu\Sc^{(0)}_{\nu\mu}=0$, which \textcolor{black}{is the constraint equation (\ref{B})
to zeroth order in $\hbar$}.
The general solution \textcolor{black}{of} the above equations reads
\begin{eqnarray}
\F^{(0)} &=& m \,V^{(0)}\, \delta(p^2-m^2)\,, \n\\
\Sc^{(0)}_{\mu\nu} &=& m\, \Sigma^{(0)}_{\mu\nu} A^{(0)}\, \delta(p^2-m^2) \,, \label{FA0}
\end{eqnarray}
where $V^{(0)}, \Sigma^{(0)}_{\mu\nu}A^{(0)}$ are up to now arbitrary functions which do not have singularities
at $p^2=m^2$. We also demand that they go to zero sufficiently fast for large momenta (in order to neglect boundary
terms when performing an integration by parts). Comparing to the previous section, we can identify $V^{(0)}$ with
the \textcolor{black}{spin-symmetric combination} (\ref{V_Wignerdirect}) and $A^{(0)}$ with 
\textcolor{black}{spin-anti-symmetric combination} (\ref{A_Wignerdirect})
of the zeroth-order distribution function, as well as $\Sigma^{(0)}_{\mu\nu}$ with the 
dipole-moment tensor,
which satisfies $p^\mu \Sigma^{(0)}_{\mu\nu}=0$ in order to fulfill Eq.\ (\ref{B}).
In order to \textcolor{black}{be consistent} with Eq.\ (\ref{Sigmadirect}), we 
demand $\Sigma^{(0)\mu\nu}\Sigma^{(0)}_{\mu\nu}=2$.

With the help of Eq.\ (\ref{PVS}) we can now write down the remaining components of the Wigner function
to leading order in $\hbar$,
\begin{eqnarray}
\Pc^{(0)}&=&0\, , \n\\
\V^{(0)}_\mu &=& p_\mu V^{(0)}\, \delta(p^2-m^2) \,, \n\\
\A^{(0)}_{\mu} &=& -\frac{1}{2}\epsilon_{\mu\nu\alpha\beta}\, p^\nu \Sigma^{(0)\alpha\beta}A^{(0)} \, \delta(p^2-m^2) \,.
\label{PVS0}
\end{eqnarray}
It is straightforward to check that our solutions (\ref{FA0}), (\ref{PVS0}) satisfy Eqs.\ (\ref{Vkin}) -- (\ref{Akin}).
All zeroth-order solutions are on mass-shell and agree with the results from the direct calculation of the Wigner function
in Sec.\ \ref{zerothorderWigner_direct}.

\subsection{First order in $\hbar$}

The starting point for our analysis of the contributions of next-to-leading order in $\hbar$ is again the on-shell
equation (\ref{on-shell}). The \textcolor{black}{$\mathcal{O}(\hbar)$} part reads
\begin{eqnarray}
(p^2-m^2)\F^{(1)}&=& \frac{1}{2}p^\mu \nabla^{(0)\nu} \Sc_{\nu\mu}^{(0)}=\frac{1}{2}F^{\mu\nu}\Sc^{(0)}_{\mu\nu}\,,\n\\
(p^2-m^2)\Sc^{(1)}_{\mu\nu} &=& -p^\alpha p_{[\mu}\Sc^{(1)}_{\nu]\alpha}-\frac{1}{2}\nabla^{(0)}_{[\mu}p_{\nu]}\F^{(0)}
 \n\\ &=& F_{\mu\nu}\F^{(0)}\,, \label{FS1storder}
\end{eqnarray}
where we used $p^\mu \Sc^{(0)}_{\nu\mu}=0$ and the relation
\begin{equation}
p^\nu\Sc^{(1)}_{\mu\nu}=\frac{1}{2}\nabla^{(0)}_\mu \F^{(0)} \,, \label{pS1}
\end{equation}
which follows from Eq.\ (\ref{B}) to first order in $\hbar$.
Here the leading-order functions $\Sc_{\mu\nu}^{(0)}$ and $\F^{(0)}$ have been obtained in the previous subsection.
The solutions to Eq.\ (\ref{FS1storder}) can in general be written as
\begin{eqnarray}
\F^{(1)} &=& m \left[V^{(1)}\,\delta(p^2-m^2)\right. \n \\
&  & \hspace*{0.3cm} -\left. \frac{1}{2}F^{\mu\nu}\Sigma_{\mu\nu}^{(0)}A^{(0)}\,\delta^{\prime}(p^2-m^2) \right]\,,  \n\\
\Sc^{(1)}_{\mu\nu} &=& m \left[\bar{\Sigma}^{(1)}_{\mu\nu}\delta(p^2-m^2)-F_{\mu\nu} V^{(0)}\delta^{\prime}(p^2-m^2)
\right]\,.\n \\
\label{FA1}
\end{eqnarray}
Here, $\bar{\Sigma}^{(1)}_{\mu\nu}$
is, up to a factor $m$, the on-shell part of the first-order dipole moment.
We note that $\bar{\Sigma}^{(1)}_{\mu\nu}$ is not normalized.
The functions $V^{(1)}$ and $\bar{\Sigma}^{(1)}_{\mu\nu}$ will be determined from the kinetic \textcolor{black}{equations}
that we will derive below. \textcolor{black}{The function} 
$V^{(1)}$ can be identified as the $\mathcal{O}(\hbar)$ correction to the
\textcolor{black}{spin-symmetric combination} of the distribution function. 
\textcolor{black}{Using} Eq.\ (\ref{pS1}), we derive a constraint for $\bar{\Sigma}^{(1)}_{\mu\nu}$,
\begin{equation}
p^\nu \bar{\Sigma}^{(1)}_{\mu\nu} \, \delta(p^2-m^2)=\frac12\, \delta(p^2-m^2)\, \nabla_\mu^{(0)}V^{(0)}\,. \label{heyomega}
\end{equation}

Expanding all quantities in Eq.\ (\ref{PVS}) into power series in $\hbar$, to $\mathcal{O}(\hbar)$ we obtain
\begin{eqnarray}
\Pc^{(1)}&=&-\frac{1}{2m}\nabla^{(0)\mu}\A^{(0)}_\mu \,, \n\\
\V^{(1)}_\mu&=&\frac{p_\mu}{m}\F^{(1)}-\frac{1}{2m}\nabla^{(0)\nu}\Sc^{(0)}_{\nu\mu}\,,  \n\\
\A^{(1)}_{\mu}&=& \frac{1}{2m}\nabla_\mu^{(0)}\Pc^{(0)}-\frac{1}{2m}\epsilon_{\mu\nu\alpha\beta}
p^\nu \Sc^{(1)\alpha\beta} \, . \label{PVA1_org}
\end{eqnarray}
Inserting the zeroth- and first-order solutions from Eqs.\ (\ref{FA0}), (\ref{PVS0}), and (\ref{FA1}), we can derive
the first-order pseudo-scalar, vector, and axial-vector functions,
\begin{eqnarray}
\Pc^{(1)}&=& \frac{1}{4m}\epsilon^{\mu\nu\alpha \beta}\nabla_\mu^{(0)}
\left[p_\nu \Sigma_{\alpha\beta}^{(0)}A^{(0)}\, \delta(p^2-m^2)\right] \,, \n\\
\V^{(1)}_\mu&=&\delta(p^2-m^2)\left[p_\mu V^{(1)}+\frac{1}{2}\nabla^{(0)\nu}\Sigma_{\mu\nu}^{(0)}A^{(0)}\right] \n\\
&-& \left[\frac{1}{2}p_\mu F^{\alpha\beta}\Sigma_{\alpha\beta}^{(0)}
+ \Sigma^{(0)}_{\mu\nu}F^{\nu\alpha}p_\alpha \right]A^{(0)}\delta^{\prime}(p^2-m^2)\,,  \n \\
\A^{(1)}_{\mu}&=&
m\,\bar{n}^{(1)}_\mu \delta(p^2-m^2)+\tilde{F}_{\mu\nu}p^\nu V^{(0)}\, \delta^\prime(p^2-m^2), \label{PVA1}
\end{eqnarray}
where
\begin{equation}
\bar{n}^{(1)}_\mu\equiv -\frac{1}{2m}\epsilon_{\mu\nu\alpha\beta}p^\nu\bar{\Sigma}^{(1)\alpha\beta} \label{def_n1}
\end{equation}
is the first-order on-shell correction to $n^{(0)}_\mu A^{(0)}$.

To first order in $\hbar$, the constraints (\ref{Vkin}), (\ref{Akin}) read
\begin{eqnarray}
\nabla^{(0)} \cdot \V^{(0)}&=&0 \,, \n\\
p_{[\mu} \V_{\nu]}^{(1)}+\frac12 \epsilon_{\mu\nu\alpha\beta}\nabla^{(0)\alpha}\A^{(0)\beta}&=&0 \,.\label{constrain1}
\end{eqnarray}
They lead to the kinetic equations of the particle distributions and the dipole moment to zeroth order in $\hbar$;
for details see App.\ \ref{appendixB},
\begin{eqnarray}
\delta(p^2-m^2)\, p\cdot\nabla^{(0)} V^{(0)}&=&0 \,,\n\\
\delta(p^2-m^2)\, p\cdot\nabla^{(0)} A^{(0)}&=&0 \,,\n\\
\delta(p^2-m^2)\left[p\cdot \nabla^{(0)} \Sigma_{\mu\nu}^{(0)}
-F_{\ [\mu}^{ \alpha}\Sigma^{(0)}_{\nu]\alpha}\right]&=&0\,. \label{kin0}
\end{eqnarray}

\subsection{Second order in $\hbar$}

As we have shown in the previous subsection, the zeroth-order kinetic equations are derived from the first-order
constraint equations. In order to obtain the first-order kinetic equations, we focus on the second-order parts of
Eqs.\ (\ref{Vkin}), (\ref{Akin}),
\begin{eqnarray}
\nabla^{(0)} \cdot \V^{(1)}&=&0\,, \n\\
p_{[\mu}\V_{\nu]}^{(2)}+\Pi^{(2)}_{[\mu}\V_{\nu]}^{(0)}
+\frac12\epsilon_{\mu\nu\alpha\beta}\nabla^{(0)\alpha}\A^{(1)\beta}&=&0\,, \label{constrain2}
\end{eqnarray}
with the operator $\Pi_\mu^{(2)}=-\frac{1}{12}(\partial_{x \alpha} F_{\mu\nu})\partial_p^\alpha \partial_p^\nu$. After some
calculation (cf.\ App.\ \ref{appendixB}), one derives the following kinetic equations,
\begin{eqnarray}
0 &=& \delta(p^2-m^2)\Big[p \cdot \nabla^{(0)}V^{(1)} \Big. \n\\
& &  +\Big. \frac{1}{4}(\partial_x^\alpha F^{\mu\nu})
\partial_{p\alpha} \left(\Sigma_{\mu\nu}^{(0)}A^{(0)}\right)\Big] \n\\
&& -\frac{1}{2}\delta^\prime(p^2-m^2)F^{\alpha\beta}p \cdot \nabla^{(0)}\left(\Sigma_{\alpha\beta}^{(0)}A^{(0)}\right)\,, \n\\
0 & = &\delta(p^2-m^2) \Big[p\cdot \nabla^{(0)}\bar{\Sigma}_{\mu\nu}^{(1)} \Big. \n \\
&& -\Big. F_{\ [\mu}^{\alpha}\bar{\Sigma}_{\nu]\alpha}^{(1)}
+\frac{1}{2} (\partial_{x\alpha} F_{\mu\nu})\partial_p^\alpha V^{(0)} \Big] \n \\
&& -\delta^\prime(p^2-m^2) F_{\mu\nu}\, p \cdot \nabla^{(0)}V^{(0)} \,.
\label{kin1}
\end{eqnarray}
\textcolor{black}{Multiplying the second equation (\ref{kin1}) by $-\frac{1}{2m} \epsilon^{\alpha \beta \mu \nu} p_\beta$
and using Eq.\ (\ref{def_n1}),
we obtain a kinetic equation for $\bar{n}_\mu^{(1)}$,}
\begin{eqnarray}
0 &=& \delta(p^2-m^2)\Big[p\cdot\nabla^{(0)}\bar{n}^{(1)}_{\mu}-F_{\mu\nu}\bar{n}^{(1)\nu}\Big. \n\\
&& -\Big.\frac{1}{2m}p^\nu(\partial_{x\alpha}\tilde{F}_{\mu\nu})\partial_p^\alpha V^{(0)}\Big] \n \\
&& + \delta^\prime(p^2-m^2) \frac1m \tilde{F}_{\mu\nu} p^\nu p \cdot \nabla^{(0)}V^{(0)}
\,,\label{kin_n1}
\end{eqnarray}
\textcolor{black}{where $\tilde{F}_{\mu \nu} \equiv \frac12 \epsilon_{\mu \nu \alpha \beta} F^{\alpha\beta}$ is
the dual field-strength tensor.}

\section{Kinetic equations for spin-1/2 particles}

In order to summarize our results in a compact form, we define
the resummed functions
\begin{eqnarray}
\label{resummedVSigma}
V & \equiv & V^{(0)}+\hbar V^{(1)}+\mathcal{O}(\hbar^{2})\, ,\nonumber \\
\bar{\Sigma}^{\mu\nu} & \equiv & \Sigma^{(0)\mu\nu}A^{(0)}+\hbar\bar{\Sigma}^{(1)\mu\nu}+\mathcal{O}(\hbar^{2})\, .
\end{eqnarray}
Using these resummed functions,
the components the Wigner function, given by Eqs.\ \eqref{FA0}, \eqref{PVS0} to zeroth order in $\hbar$
and by Eqs.\ \eqref{FA1}, \eqref{PVA1} to first order in $\hbar$,
can be written as 
\begin{eqnarray}
\mathcal{F} & = & m\left[V\,\delta(p^{2}-m^{2})-\frac{\hbar}{2}F^{\mu\nu}\bar{\Sigma}_{\mu\nu}\,\delta^{\prime}(p^{2}-m^{2})\right]\n\\
&&+\mathcal{O}(\hbar^{2})\, ,\nonumber \\
\mathcal{P} & = & \frac{\hbar}{4m}\epsilon^{\mu\nu\alpha\beta}\nabla_{\mu}^{(0)}\left[p_{\nu}\bar{\Sigma}_{\alpha\beta}\,\delta(p^{2}-m^{2})\right]+\mathcal{O}(\hbar^{2})\, ,\nonumber \\
\mathcal{V}_{\mu} & = & p_{\mu}\left[V\,\delta(p^{2}-m^{2})-\frac{\hbar}{2}F^{\alpha\beta}\bar{\Sigma}_{\alpha\beta}\,\delta^{\prime}(p^{2}-m^{2})\right]\nonumber \\
 &  & +\frac{\hbar}{2}\nabla^{(0)\nu}\left[\bar{\Sigma}_{\mu\nu}\,\delta(p^{2}-m^{2})\right]+\mathcal{O}(\hbar^{2})\, ,\nonumber \\
\mathcal{A}_{\mu} & = & -\frac{1}{2}\epsilon_{\mu\nu\alpha\beta}p^{\nu}\left[\bar{\Sigma}^{\alpha\beta}\,\delta(p^{2}-m^{2})\right.\n\\
&&-\left.\hbar F^{\alpha\beta}V\,\delta^{\prime}(p^{2}-m^{2})\right]+\mathcal{O}(\hbar^{2})\, ,\nonumber \\
\mathcal{S}_{\mu\nu} & = & m\left[\bar{\Sigma}_{\mu\nu}\,\delta(p^{2}-m^{2})-\hbar F_{\mu\nu}V\,\delta^{\prime}(p^{2}-m^{2})\right]\n\\
&&+\mathcal{O}(\hbar^{2})\, .\label{eq:resummed-Wigner}
\end{eqnarray}
The undetermined functions
$V$ and $\bar{\Sigma}_{\mu\nu}$ satisfy one constraint equation,
\begin{equation} \label{eq:constraintt}
p^{\nu}\bar{\Sigma}_{\mu\nu}\, \delta(p^2-m^2)=\frac{\hbar}{2}\, \delta(p^2-m^2)\,
\nabla_{\mu}^{(0)}V+\mathcal{O}(\hbar^{2})\,,
\end{equation}
and two kinetic equations, which are the sum of Eqs.~\eqref{kin0} and \eqref{kin1},
\begin{eqnarray}
0 & = & \delta(p^{2}-m^{2})\left[p\cdot\nabla^{(0)}V+\frac{\hbar}{4}(\partial_{x}^{\alpha}F^{\mu\nu})\partial_{p\alpha}\bar{\Sigma}_{\mu\nu}\right]\nonumber \\
 &  & -\frac{\hbar}{2}\delta^{\prime}(p^{2}-m^{2})F^{\alpha\beta}\,p\cdot\nabla^{(0)}\bar{\Sigma}_{\alpha\beta}+\mathcal{O}(\hbar^{2})\, ,\nonumber \\
0 & = & \delta(p^{2}-m^{2})\bigg[p\cdot\nabla^{(0)}\bar{\Sigma}_{\mu\nu}-F_{\ [\mu}^{\alpha}\bar{\Sigma}_{\nu]\alpha}\bigg.\n\\
&&\bigg.+\frac{\hbar}{2}(\partial_{x\alpha}F_{\mu\nu})\partial_{p}^{\alpha}V\bigg]\nonumber \\
 &  & -\hbar\delta^{\prime}(p^{2}-m^{2})F_{\mu\nu}\,p\cdot\nabla^{(0)}V+\mathcal{O}(\hbar^{2})\, .\label{eq:kin-VandSigmaA}
\end{eqnarray}

Up to first order, we find that Eqs.~\eqref{eq:resummed-Wigner}, \eqref{eq:constraintt}, and \eqref{eq:kin-VandSigmaA} are
invariant under the following transformation
\begin{eqnarray}
\bar{\Sigma}_{\mu\nu} & \rightarrow &\widehat{\bar\Sigma}_{\mu\nu}=\bar{\Sigma}_{\mu\nu}+(p^{2}-m^{2})\delta\bar{\Sigma}_{\mu\nu}\, ,\nonumber \\
V & \rightarrow & \widehat{V}= V-\frac{\hbar}{2}F^{\mu\nu}\delta\bar{\Sigma}_{\mu\nu}\,, \label{eq:off-shell-trans-1}
\end{eqnarray}
or the transformation
\begin{eqnarray}
V & \rightarrow & \widehat{V}=V+(p^{2}-m^{2})\delta V\, ,\nonumber \\
\bar{\Sigma}_{\mu\nu} & \rightarrow & \widehat{\bar\Sigma}_{\mu\nu}=\bar{\Sigma}_{\mu\nu}-\hbar F_{\mu\nu}\delta V\, .\label{eq:off-shell-trans-2}
\end{eqnarray}
Here $\delta\bar{\Sigma}_{\mu\nu}$ and $\delta V$ are arbitrary functions,
which should be nonsingular on the mass-shell $p^{2}=m^{2}$. The
invariance can be easily proved by using the property of the Dirac $\delta$-function
$-x\delta^{\prime}(x)=\delta(x)$. 
Note that the first (second) transformation
does not affect the on-shell value of $\bar{\Sigma}_{\mu\nu}$ ($V$) because
the factor $p^{2}-m^{2}$ in front of $\delta \bar{\Sigma}_{\mu \nu}$ ($\delta V$) vanishes on the
mass-shell. 

It is possible to show that without loss of generality one can omit the terms proportional to the derivative of the delta 
function in the kinetic equations \eqref{eq:kin-VandSigmaA}. In order to prove this, let us consider
the $p^0$-integrated version of the last term in the second kinetic equation
\eqref{eq:kin-VandSigmaA}. For any function $G(x,p)$, we have
\begin{eqnarray}
 &&\int dp^0\, \delta'(p^2-m^2)G(x,p)p\cdot\nabla^{(0)}V\n\\
 &=&\int dp^0\, \frac{1}{2p^0}G(x,p)\left[\partial_{p^0}\delta(p^2-m^2)\right]p\cdot\nabla^{(0)}V\n\\
 &=&-\int dp^0\, \frac{1}{2p^0}\delta(p^2-m^2)G(x,p)\partial_{p^0}\, p\cdot\nabla^{(0)}V+\mathcal{O}(\hbar)\, ,\n\\ 
 \label{deltaprimeterm}
\end{eqnarray}
where we integrated by parts in the last step and used Eq.\ \eqref{kin0}. Applying the transformation 
\eqref{eq:off-shell-trans-2} to Eq.\ \eqref{deltaprimeterm} and choosing $\delta V$ such that
\begin{eqnarray}
 \delta(p^2-m^2)2p^{0} p\cdot\nabla^{(0)}\delta V&=&-\delta(p^2-m^2)\partial_{p^0} p\cdot \nabla^{(0)}V\, ,\n \\
 \label{deltaSigma_condition}
\end{eqnarray}
(where we assume that $\delta V$ is non-singular at $p^2 = m^2$)
we find
\begin{eqnarray}
 \int dp^0\, \delta'(p^2-m^2)G(x,p)p\cdot\nabla^{(0)}\widehat{V}=\mathcal{O}(\hbar)\, .
\end{eqnarray}
A similar procedure can be applied to the first kinetic equation \eqref{eq:kin-VandSigmaA}.
This proves that the terms proportional to the derivative of the delta function in the kinetic equations
\eqref{eq:kin-VandSigmaA} are actually of order $O(\hbar^2)$, and we obtain
%
\begin{eqnarray}
0 & = & \delta(p^{2}-m^{2})\left[ p\cdot\nabla^{(0)}\widehat{V}+\frac{\hbar}{4}(\partial_{x}^{\alpha}F^{\mu\nu})\partial_{p\alpha}\widehat{\bar{\Sigma}}_{\mu\nu}\right] \nonumber\\
&&+\mathcal{O}(\hbar^{2})\, ,\nonumber \\
0 & = & \delta(p^{2}-m^{2})\bigg[ p\cdot\nabla^{(0)}\widehat{\bar{\Sigma}}_{\mu\nu}-F_{\ [\mu}^{\alpha}\widehat{\bar{\Sigma}}_{\nu]\alpha} \bigg. \nonumber\\
&&\bigg. +\frac{\hbar}{2}(\partial_{x\alpha}F_{\mu\nu})\partial_{p}^{\alpha}\widehat{V} \bigg]+\mathcal{O}(\hbar^{2})\, . \label{eq:kin-after-transformation}
\end{eqnarray}
The kinetic equations (\ref{eq:kin-after-transformation}) are the main result of the present paper. For the sake of 
notational convenience, we will omit the hat in the following. 

In order to write the first kinetic equation (\ref{eq:kin-after-transformation}) in terms of the distribution functions, we define
\begin{equation}
V(x,p) \equiv \frac{2}{(2\pi \hbar)^3}\sum_{es} \theta(e\,p^0)f^{e}_s(x,e\,\mathbf{p}) \label{V_Wignerdirect11} \, ,
\end{equation}
where $f_s^{\pm}=f_s^{(0)\pm}+\hbar f_s^{(1)\pm}$.
Because of the theta function, the support of the distribution function for anti-particles is different from the
one for particles. Thus, these distribution functions have to fulfill the first equation (\ref{eq:kin-after-transformation}) 
separately \cite{Vasak:1987um}. Then, using Eqs.~\eqref{A_Wignerdirect}, \eqref{resummedVSigma}, 
and \eqref{V_Wignerdirect11}, the first equation \eqref{eq:kin-after-transformation} can be written as
\begin{eqnarray}
0 & =  &\sum_s\delta\left(p^2-m^2\right)\left[p \cdot \nabla^{(0)} \right. \n \\
&  & \hspace*{1cm} + \left. s \frac{\hbar}{4} (\partial_x^{\mu}F^{\nu\rho})
\partial_{p\mu}  \Sigma_{\nu\rho}^{(0)} \right] \theta(\pm p_0)f^\pm_{ s}\, .
\label{GBE}
\end{eqnarray}

To conclude this section, we remark that the terms containing the derivative of the delta function, 
although they do not contribute to the kinetic equations, lead to a modification of the on-shell condition of 
the components of the Wigner function. Noting that
\begin{eqnarray}
\lefteqn{\delta\left(p^2-m^2-s \frac{\hbar}{2} F^{\mu\nu}\Sigma_{\mu\nu}^{(0)}\right) =\delta(p^2-m^2) }\n \\
 & & \hspace*{1.7cm} -s \frac{\hbar}{2} F^{\mu\nu}\Sigma_{\mu\nu}^{(0)}\delta'(p^2-m^2)+\mathcal{O}(\hbar^2)\,,
\end{eqnarray}
we can for instance combine Eqs.\ (\ref{FA0}) and (\ref{FA1}) and use Eqs.\ (\ref{V_Wignerdirect}) and
(\ref{A_Wignerdirect}) to obtain to order $\mathcal{O}(\hbar)$
\begin{eqnarray}
 \F &=& \F^{(0)} + \hbar \F^{(1)}\n \\
 & = & \frac{2}{(2\pi \hbar)^3}m\sum_s \delta \left(p^2-m^2-s \frac{\hbar}{2} F^{\mu\nu}\Sigma_{\mu\nu}^{(0)}\right)\n \\ &&
\times \left[\theta(p^0)f^+_s+\theta(-p^0)f^-_{s}\right]\,.
\end{eqnarray}
Thus, to first order in $\hbar$ the on-shell condition is modified to
\begin{equation}
p^2=m_s^2\equiv m^2+ s \frac{\hbar}{2} F^{\mu\nu}\Sigma_{\mu\nu}^{(0)}\,.
\end{equation}
In the following, we discuss the massless limit and
the classical case, as well as some consequences for global equilibrium and fluid dynamics.

\section{Massless limit}

In this section, we explain how to obtain the massless limit of the currents $\V_\mu$ and $\A_\mu$, cf.\
Eqs.\ (\ref{PVS0}) and (\ref{PVA1}). The crucial step is to replace the dipole-moment tensor (\ref{Sigmadirect})
for $m \neq 0$ by the corresponding one for $m=0$. Obviously, this cannot \textcolor{black}{be achieved simply} by taking
the limit $m\rightarrow 0$ in Eq.\ (\ref{Sigmadirect}).

For massive particles, the dipole-moment tensor as well as the particle's position are uniquely defined in the rest frame.
The Pauli--Lubanski operator is defined as \cite{itzykson2012quantum}
\begin{equation}
\hat{N}^\mu=-\frac{1}{2m}\epsilon^{\mu\nu\rho\sigma}\sigma_{\nu\rho}\hat{P}_\sigma\,,
\end{equation}
where $\hat{P}^\mu \equiv i \hbar D^\mu$ is the (kinetic) momentum operator. In the rest frame,
the Pauli--Lubanski operator fulfills the commutation relations of an angular momentum. Let $\psi, \bar{\psi}$ be
solutions of the Dirac equation (\ref{DiracEq}). Then the dipole-moment tensor
$\Sigma^{\mu\nu}\equiv \bar{\psi}\sigma^{\mu\nu}\psi$ fulfills $p_\mu\Sigma^{\mu\nu}=0$,
where $p_\mu$ is the eigenvalue of $\hat{P}_\mu$. Thus,
\begin{equation}
\Sigma^{\mu\nu}=-\frac1m\epsilon^{\mu\nu\alpha\beta}p_\alpha n_\beta \,, \label{Sigma}
\end{equation}
with $n_\beta=\bar{\psi}\hat{N}_\beta\psi$. This agrees with Eq.\ (\ref{Sigmadirect}), if $\psi = u$ or $v$.

On the classical level, $\Sigma^{\mu\nu}$ is the intrinsic angular-momentum tensor about the center
of mass. In a relativistic theory, the center of mass of a particle is frame-dependent. In order to have a
frame-independent definition of $\Sigma^{\mu \nu}$, one requires
$p_\mu \Sigma^{\mu\nu}=0$ as a gauge condition. This requirement identifies the dipole-moment tensor
(\ref{Sigma}) as the intrinsic angular-momentum tensor about the center of mass in the rest frame
of the particle \cite{Stone:2014fja}.

For massless particles there is no rest frame, thus both the position (in the classical case the center of momentum)
and the dipole-moment tensor can at first be defined in an arbitrary frame, which makes them frame-dependent.
For massless particles, the polarization vector $n^\mu$ is always parallel to the momentum $p^\mu$. Thus,
the requirement $p_\mu\Sigma^{\mu\nu}=0$ can no longer be used as a gauge condition, since Eq.\ (\ref{Sigma})
automatically satisfies this constraint. [In the massless limit, one also needs to change the normalization
of the spinors to $\bar{u}u=2\vert \mathbf{p} \vert$ \cite{DeGroot:1980dk}.]
If we choose the dipole-moment tensor to be defined
in a frame characterized by a time-like four-vector $u^\mu$, we can choose the gauge condition
$u^\mu \Sigma_{\mu\nu}=0$ \cite{Chen:2015gta}.
Consequently, the frame vector $u^\mu$ must assume the role of $p^\mu$ in Eq.\ (\ref{Sigma}).
Moreover, since $n^\mu$ and $p^\mu$ are parallel for massless particles, the momentum $p^\mu$
can assume the role of $n^\mu$ in Eq.\ (\ref{Sigma}). Finally, in order to obtain the massless case
we need to replace the normalization factor $1/m$ in Eq.\ (\ref{Sigma}).
The energy of a massive particle in its rest frame is $p^0_{rf} = \sqrt{p^2}$. If the particle is on the mass-shell,
this is equivalent to $p^0_{rf}=m$.
The energy of a massless particle in the rest frame of $u^\mu$, however, is $p^0_u = p \cdot u$.
Thus, it is natural to replace the normalization $1/m$ in Eq.\ (\ref{Sigma}) by $1/(p\cdot u)$. We emphasize that this
replacement can only be done in the presence of a $\delta$-function which sets the rest-frame energy equal to the
mass $m$.
The explicit expression for the dipole-moment tensor in the massless case \textcolor{black}{is then} given by
\begin{equation}
\Sigma^{\mu\nu}_u=- \frac{1}{p\cdot u}\epsilon^{\mu\nu\alpha\beta}u_\alpha p_\beta\,, \label{SigmaMassless}
\end{equation}
which agrees with the definition of the ``spin tensor'' in Ref.\ \cite{Chen:2015gta}.
This tensor corresponds classically to the intrinsic angular momentum about the center of momentum
as seen from the frame where $u^\mu=(1,0,0,0)$ and will have the quantum-mechanical properties of an
angular-momentum operator in that frame.

With this knowledge, we can make the transition between the Wigner functions of massive and massless particles.
For zero fermion mass, Eqs.\ (\ref{S}) and (\ref{Akin}) decouple. By defining right- and left-handed
currents $J^\mu_\chi \equiv \frac12 (\V^\mu + \chi\A^\mu)$, $\chi=\pm$ \textcolor{black}{for right-/left-handed particles}, 
we have to order $\hbar$
\begin{equation}
\frac{\hbar}{2} \left(\nabla^\mu J^\nu_\chi-\nabla^\nu J^\mu_\chi\right)=\chi\,\epsilon^{\mu\nu\alpha\beta}
p_\alpha J_{\chi,\beta}\,.
\end{equation}
These equations have been solved in Refs.\ \cite{Hidaka:2016yjf,Huang:2018wdl,Gao:2018wmr}, with the result
\begin{eqnarray}
J^\mu_\chi&=&\delta(p^2) \left( p^\mu + \chi\, \frac{\hbar}{2}\, \Sigma^{\mu \nu}_u \nabla_\nu\right) f_\chi
\n \\
&& + \chi\, \hbar \tilde{F}^{\mu \nu} p_\nu \delta'(p^2)f_\chi \,, \label{Hidaka}
\end{eqnarray}
where $f_\chi$ is the distribution function for right-/left-handed fermions and $u^\mu$ is the four-velocity of an
arbitrary frame.
We remark that in the massive case, $s$ describes ``spin up'' or ``spin down'', which corresponds to positive or negative 
helicity in the massless limit ($mn_\mu\rightarrow p_\mu$). On the other hand, the currents above are defined for given 
chirality $\chi$. Since helicity and chirality are identical for massless particles, but opposite for massless anti-particles, 
the relation between chirality $\chi$ and spin/helicity $s$ is $\chi = e\,s$ with $e=\pm$ representing particles/anti-particles.

To obtain the massless limit of our solutions, we replace the massive dipole-moment tensor by the massless one,
$\Sigma^{(0)\mu\nu}\rightarrow \Sigma^{\mu\nu}_u$. In order to obtain the vector current for the massless case from
Eq.\ (\ref{PVA1}), we need to consider the term $\sim \nabla^{(0)\nu} \Sigma_{\mu\nu}^{(0)}$. We first pull
the constant factor $1/m$ out of the derivative and then replace
$\delta(p^2-m^2)/m=\delta(p^2-m^2)/\sqrt{p^2}\rightarrow \delta(p^2)/(p\cdot u)$.
Finally, replacing $p^\mu/m \rightarrow u^\mu$, $m n^\mu \rightarrow p^\mu$
in this term we find
\begin{eqnarray}
 \V^{(1)}_{\mu,m=0}&=&\delta(p^2)\left[p_\mu V^{(1)}+\frac{1}{2\,p\cdot u}\epsilon_{\mu\nu\alpha\beta}
 \nabla^{(0)\nu}p^\alpha u^\beta A^{(0)}\right]\n \\
 &-&  \left[\frac{1}{2}p_\mu F^{\alpha\beta}\Sigma_{u, \alpha\beta}
+ \Sigma_{u,\mu\nu}F^{\nu\alpha}p_\alpha \right]A^{(0)}\delta^{\prime}(p^2)\,. \n \\ \label{Vm0}
\end{eqnarray}
In Ref.\ \cite{Hidaka:2016yjf} the frame-vector $u_\mu$ is assumed to be independent of space-time coordinates.
In order to compare to the solution found in that reference, we adopt the same assumption. Evaluating the derivatives,
contracting the $\epsilon$-tensors, and using $p^2\delta'(p^2)=- \delta(p^2)$, we find 
\textcolor{black}{from Eqs.\ (\ref{PVS0}) and (\ref{Vm0})}
\begin{eqnarray}
 \mathcal{V}^\mu_{m = 0}&=& \delta(p^2)\left[p^\mu V+\frac{\hbar}{2}\Sigma_u^{\mu \nu} \nabla_\nu^{(0)} A^{(0)}\right]\n\\
 && +\hbar\tilde{F}^{\mu\nu}p_\nu A^{(0)}\delta'(p^2)+\mathcal{O}(\hbar^2)\, , \label{Vm0_final} 
\end{eqnarray}
\textcolor{black}{where} $V=V^{(0)}+\hbar V^{(1)}$. Note that $V^{(1)}$ depends on the frame vector 
$u^\mu$ such that the whole expression \eqref{Vm0_final} is frame independent 
\cite{Chen:2015gta,Huang:2018wdl,Gao:2018jsi}.
To obtain the axial-vector current in the massless case from \textcolor{black}{Eqs.\ (\ref{PVS0}) and} (\ref{PVA1}),
we note that the general solution of Eq.\ \eqref{heyomega} reads
\begin{equation}
\bar{\Sigma}^{(1)}_{\mu\nu}=\Sigma_{v,\mu\nu}A^{(1)}+\frac{1}{2\,p\cdot u}
\left(u_\nu \nabla_\mu^{(0)}-u_\mu \nabla_\nu^{(0)}\right)V^{(0)}\,, \label{chi_massless}
\end{equation}
where the first and second terms depend on  arbitrary time-like unit vectors $u^\mu$ and $v^\mu$, respectively. 
Here, one makes use of the first equation (\ref{kin0}) to see that
the constraint (\ref{heyomega}) is fulfilled. Inserting Eq.\ (\ref{chi_massless}) into Eq.\ (\ref{PVA1}), 
and replacing the zeroth order dipole-moment tensor $\Sigma^{(0)}_{\mu\nu}$ by $\Sigma_{u,\mu\nu}$, we find
\begin{eqnarray}
 \mathcal{A}^\mu_{m=0}&=& \delta(p^2) \left[ p^\mu A
 +\frac{\hbar}{2} \Sigma^{\mu\nu}_u \nabla_\nu^{(0)}V^{(0)} \right]\n\\
 &&  +\hbar\tilde{F}^{\mu\nu}p_\nu V^{(0)}\delta'(p^2)+\mathcal{O}(\hbar^2)\,, \label{Am0}
\end{eqnarray}
where $A\equiv A^{(0)}+\hbar A^{(1)}$, with $A^{(1)}$ dependent on $u^\mu$. Note that, in order for the 
\textcolor{black}{above axial current} to be frame-independent, 
the function $A^{(1)}$ cannot depend on $v^\mu$. Adding and 
subtracting Eqs.\ (\ref{Vm0_final}) and (\ref{Am0}), we recover the result (\ref{Hidaka}). 
Acting with $\nabla_\mu$ on this equation, one can derive the chiral kinetic theory of Refs.\
\cite{Hidaka:2016yjf,Hidaka:2017auj,Huang:2018wdl,Gao:2018wmr,Yang:2018lew,Gao:2018jsi}.

\section{Comparison to the classical case}

In this section, we show that Eq.\ (\ref{GBE}) gives rise to the first and second Mathisson--Papapetrou--Dixon (MPD)
equations  \cite{Bailey:1975fe,Israel:1978up} as well as to the BMT equation  \cite{Bargmann:1959gz}, which
were derived for classical, extended, spinning particles with non-vanishing dipole moment.
Comparing Eq.\ (\ref{GBE}) to the generic form of the collisionless relativistic Boltzmann--Vlasov 
equation~\cite{Israel:1978up,Cercignani}
\begin{equation}
 p \cdot \partial_{x}f_s+m\,\partial_{p\mu}(F_s^\mu f_s)=0\,, \label{colllessBVE}
\end{equation}
where $f_s$ is the distribution function, $F_s^\mu=d p^\mu/d\tau$ is the external force, 
$p^\mu=m\, d x^\mu /d \tau$ and $\tau$ the world-line parameter, we find that in our case
\begin{equation} \label{gen_force}
 F_s^\mu=\frac1m \left[F^{\mu\nu}p_\nu+s\frac\hbar4(\partial^\mu_x F^{\nu\rho})\Sigma_{\nu\rho}^{(0)}\right]\,,
\end{equation}
i.e., the external force is given as the sum of the Lorentz force and the Mathisson force.
This is the first MPD equation \cite{Bailey:1975fe,Israel:1978up}.
In Refs.\ \cite{Bailey:1975fe,Israel:1978up}, the kinetic equation for particles with classical dipole moment
$m^{\mu \nu}$ was derived. Our results agree with those, setting
\begin{equation}
 m_{\mu\nu}\longrightarrow g \mu_B \frac{s}{2}\Sigma_{\mu\nu}^{(0)}\,,
\end{equation}
with Bohr's magneton $\mu_B \equiv \mathfrak{e} \hbar/(2 m)$, where $\mathfrak{e}$ is the electric charge, 
and the gyromagnetic ratio $g=2$, as expected
for Dirac particles with spin 1/2.

The evolution of the dipole-moment tensor is given by the third equation (\ref{kin0}), which can be rewritten as
\begin{equation}
 m\dot{\Sigma}^{(0)}_{\mu\nu}=F^{\alpha}_{\ [\mu}\Sigma^{(0)}_{\nu]\alpha}\,, \label{MPD2}
\end{equation}
where we used
\begin{equation}
 \dot{\Sigma}^{(0)}_{\mu\nu}\equiv (\dot{x}^\alpha\partial_{x\alpha}+\dot{p}^\alpha\partial_{p\alpha})\Sigma^{(0)}_{\mu\nu}
\end{equation}
with $F^\mu_{\textcolor{black}{s}}$ given by Eq.\ (\ref{gen_force}) to zeroth order. 
Equation (\ref{MPD2}) is identical to the second MPD equation \cite{Bailey:1975fe,Israel:1978up}.
Using Eq.\ (\ref{Sigmadirect}), we obtain
\begin{equation}
 m\dot{\Sigma}^{(0)}_{\mu\nu}
 =-\epsilon_{\mu\nu\alpha\beta}\left(p^\alpha\dot{n}^{(0)\beta}-\frac1mF^{\lambda\alpha}p_\lambda n^{(0)\beta}\right)\,.
\end{equation}
Inserting Eq.\ (\ref{MPD2}) and contracting with $\epsilon^{\rho\sigma\mu\nu}$ yields
\begin{equation}
 p_\rho\left(m\dot{n}^{(0)}_\sigma+F^{\mu}_{\ \sigma}n_\mu^{(0)}\right)-p_\sigma\left(m\dot{n}^{(0)}_\rho
 +F^\mu_{\ \rho} n^{(0)}_\mu\right)=0\,.
\end{equation}
Contracting with $p^\rho$ and using Eq.\ (\ref{gen_force}) to
zeroth order in $\hbar$, we conclude that
\begin{equation}
 m\dot{n}^{(0)}_\mu=F_{\mu\nu}n^{(0)\nu}\,.
\end{equation}
This is the BMT equation for classical spin precession in an electromagnetic field \cite{Bargmann:1959gz}.

\section{Global equilibrium}

Equation (\ref{GBE}) determines the single-particle distribution function $f_s^\pm$ in a general non-equilibrium
state. A special solution is obtained in global equilibrium, which we will consider in this section. 

A necessary condition for equilibrium is vanishing entropy production. Assuming the standard form of the
collision term, the distribution function in equilibrium must have the form \cite{Israel:1978up,Chen:2015gta}
\begin{equation}
f^{eq}_s=(e^{g_s}+1)^{-1}\,, \label{feq}
\end{equation}
with $g_s$ being a linear combination of the collisional invariants, namely, charge, kinetic momentum $p^\mu$,
and total angular momentum
\begin{equation}
\label{jtot}
J_s^{\mu \nu} = L^{\mu \nu} + s \frac\hbar2 \Sigma^{(0)\mu \nu} + \mathcal{O}(\hbar^2)\, ,
\end{equation}
which is the
sum of orbital angular momentum $L^{\mu \nu}= x^{[\mu} p^{\nu]}$ and spin angular momentum, 
which to first order is given by the dipole-moment tensor
$s \frac\hbar2 \Sigma^{(0)\mu \nu}$.
(Also the canonical momentum $\pi^\mu$ is conserved in a collision and could be used instead
of the kinetic momentum.
Here, we will at first use the kinetic momentum, since it is independent of space-time coordinates,
as well as gauge-independent.) Thus,
\begin{equation}
\label{ggs}
 g_s=p\cdot b(x) +a_s(x)+\frac12\Omega_{\mu\nu}(x)J_s^{\mu\nu}\,.
\end{equation}
Here, $b_\mu(x)$, $a_s(x)$, and $\Omega_{\mu\nu}(x)$ are Lagrangian multipliers, which can
depend on $x$. Since $J_s^{\mu\nu}$ is anti-symmetric, the symmetric part of $\Omega_{\mu\nu}$ can be dropped
without loss of generality.

Let us consider the case of global equilibrium with rigid rotation. Using Eq.\ \eqref{jtot}, Eq.\ \eqref{ggs} can be written as
\begin{equation}
 g_s = p\cdot \beta(x)+a_s(x)+s\frac\hbar4\Omega_{\mu\nu}(x)\Sigma^{(0)\mu\nu}\, ,
\end{equation}
where $\beta_\mu (x) \equiv b_\mu(x)+\Omega_{\nu\mu}(x)x^\nu$.
In global equilibrium, the Boltzmann equation (\ref{GBE}) needs to be fulfilled.
From the part of Eq.\ (\ref{GBE}) proportional to the derivative of $f_s^{eq}$ we obtain
\begin{eqnarray}
 0&=&\left\{p^\mu\partial_{x\mu}+\left[F^{\mu\nu}p_\nu
 +s\frac\hbar4\Sigma^{(0)}_{\nu\lambda}\left(\partial^\mu_x F^{\nu\lambda}\right)\right] \partial_{p\mu}\right\}g_s\n\\
 &=& p^\mu\left[\partial_{x\mu}a_s(x)+F_{\nu\mu}\beta^\nu(x) \right]+ p^\mu p^\nu \partial_{x\mu}\beta_\nu (x) \n\\
 &&+s\frac\hbar4 \Sigma^{(0)\rho\sigma}p \cdot\partial_{x}\Omega_{\rho\sigma}(x)
 \label{eqcond}\\
 &&+s\frac\hbar4 \left[\Sigma^{(0)\rho\sigma}\beta(x) \cdot \partial_{x}F_{\rho\sigma} 
 + \Sigma^{(0)\mu [\sigma} F^{\rho]}_{\ \ \mu}\,
\Omega_{\rho\sigma} (x)\right]\,,\n
\end{eqnarray}
where we used Eq.\ (\ref{kin0}). This equation is fulfilled, if
\begin{eqnarray}
 \partial_{x\mu} \beta_\nu + \partial_{x\nu} \beta_\mu &=&0\, , \n\\
 \partial_{x\mu} a_s(x)&=&F_{\mu\nu}\beta^\nu(x)\,,\label{chempot} \n \\
 \partial_{x\mu}\Omega_{\lambda\nu}(x)&=&0\,,
\end{eqnarray}
which makes the terms in the first and second line of Eq.\ \eqref{chempot} vanish. 
The terms in the third line of Eq.\ (\ref{eqcond}) \textcolor{black}{can be shown to vanish if
$b_\mu$ is constant,} since then $\Omega_{\mu\nu}$ is equal to the thermal vorticity, i.e.,
\begin{equation}
\Omega_{\mu\nu}=\omega_{\mu\nu}\equiv \frac12 (\partial_{x\mu} \beta_\nu - \partial_{x\nu} \beta_\mu)\, .
\end{equation}
\textcolor{black}{For the proof, one also employs the} relation
\begin{equation}
\label{relhommax}
 \beta^\alpha\partial_{x\alpha}F_{\mu\nu}-F_{\alpha\mu}\partial_{x\nu}\beta^\alpha
+F_{\alpha\nu}\partial_{x\mu}\beta^\alpha=0 \, ,
\end{equation}
which can be proven with the help of the homogeneous Maxwell equations and Eq.\ (\ref{chempot}).
These equilibrium conditions agree \textcolor{black}{with} those found in the classical case \cite{Israel:1978up} 
and those using covariant statistical mechanics~\cite{Becattini:2012tc}.
Note that the second equation (\ref{chempot}) implies that, in the rest frame of $\beta^\mu$,
an electric field is cancelled by a gradient in $a_s$.
It is amusing to note that, without electromagnetic fields, the tensor $\Omega_{\mu\nu}$ does not need to be equal 
to the thermal vorticity.

We introduce the Lie derivative of $\mathbb{A}_\mu$ along the direction of $\beta^\lambda$ as
\begin{equation}
\mathcal{L}_\beta \mathbb{A}_\mu(x) \equiv \beta(x) \cdot \partial_{x} \mathbb{A}_\mu (x)
- \mathbb{A} (x) \cdot \partial_{x} \beta_\mu(x)\,.
\end{equation}
Choosing a gauge in which $\mathcal{L}_\beta \mathbb{A}_\mu=0$, we can rewrite Eq.\ (\ref{chempot}) as
\begin{equation}
 \partial_{x\mu}\left[a_s(x)-\mathbb{A} (x)\cdot \beta(x)\right]=0\,.
\end{equation}
Defining
\begin{equation}
 -\beta(x) {\mu_s}(x) \equiv a_s(x)-\mathbb{A} (x) \cdot \beta (x)=\text{const}\,,
\end{equation}
the function $g_s$ becomes
\begin{equation}
g_s=\beta \pi  \cdot U-\beta\mu_s + s\frac\hbar4\Sigma^{(0)\mu\nu}\omega_{\mu\nu}\,.
\end{equation}
Here, $\pi_\mu\equiv p_\mu+\mathbb{A}_\mu$ is the canonical momentum, $U^\mu$ is the fluid velocity,
$\beta\equiv1/T$ is the inverse temperature, and $\mu_s$ is the chemical potential for particles with spin $s$
(for anti-particles, we need to reverse the sign of the chemical potential).
This form of $g_s$, and thus the distribution function $f_s^{eq}$, agrees in the massless and field-free limit to the
one suggested in Ref.\ \cite{Chen:2015gta}. Moreover, recalling the \textcolor{black}{definition \eqref{sigg1}, \eqref{sigg2}
of the dipole-moment tensor} one can prove that the distribution function agrees \textcolor{black}{with}
the one proposed in Ref.\ \cite{Becattini:2013fla} \textcolor{black}{to} first order in $\hbar$ if $\mu_+ =\mu_-$ 
and \textcolor{black}{if the electromagnetic field vanishes}. 

The part of Eq.\ (\ref{GBE}) which is proportional to $f_s^\pm$ vanishes if $\mu_{+}=\mu_{-}$ 
to zeroth order in $\hbar$.
In the presence of a spin imbalance, $\Delta \mu \equiv \mu_+-\mu_- \neq 0 $, it only vanishes if
\begin{equation}
(\partial_x^\lambda F^{\nu\rho})(\partial_{p\lambda}\Sigma^{(0)}_{\nu\rho})=0\,. \label{F_hom}
\end{equation}
The reason that global equilibrium with spin imbalance can in general not be realized for massive
particles is that in this case the axial-vector current is only conserved if the pseudo-scalar function
$\mathcal{P} =0$, see also Eq.\ (\ref{P}). 

To zeroth order in $\hbar$, the distribution function is given by
\begin{equation}
f^{(0)}_s=(e^{g^{(0)}_s}+1)^{-1}\,, \label{feq0}
\end{equation}
with
\begin{equation}
g^{(0)}_s=\beta(\pi\cdot U-\mu_s)\,.
\end{equation}
We define the dual thermal vorticity tensor as
$\tilde{\omega}_{\mu\nu}\equiv \frac12\epsilon_{\mu\nu\alpha\beta}\omega^{\alpha\beta}$.
Now we calculate the vector current by inserting the distribution function (\ref{feq}) into the
equation for $\V_\mu = \V^{(0)}_\mu + \hbar \V^{(1)}_\mu$, cf.\ Eqs.\ (\ref{PVS0}), (\ref{PVA1}). With
\begin{eqnarray}
 \nabla^{(0)}_\nu f_s^{(0)}&=&\frac{\partial f_s^{(0)}}{\partial g_s^{(0)}}(\pi^\alpha \partial_{x\nu} \beta_\alpha
 +\beta^\alpha \partial_{x\nu} \mathbb{A}_\alpha-F_{\nu\alpha}\beta^\alpha\n)\\
 &=&\frac{\partial f_s^{(0)}}{\partial g_s^{(0)}} p^\alpha\omega_{\nu\alpha}\,,
 \label{ooomega}
\end{eqnarray}
where we used $\mathcal{L}_\beta \mathbb{A}_\alpha=0$, Taylor-expanding
\begin{equation}
 f_s^{eq}=f_s^{(0)}+\frac{s\hbar}{4} \Sigma^{(0)}_{\mu\nu}\omega^{\mu\nu}\frac{\partial f_s^{(0)}}{\partial g_s^{(0)}}
 +\mathcal{O}(\hbar^2)\,,
\end{equation}
and noting that
\begin{eqnarray}
&& \hspace*{-1cm} \delta(p^2-m^2)\left(\Sigma^{(0)\mu\nu}p^\rho\omega_{\nu\rho}
+\frac12\Sigma^{(0)\rho\nu}p^\mu\omega_{\rho\nu}\right)\n \\
 &=& -\delta(p^2-m^2) m\tilde{\omega}^{\mu\nu}n_\nu\,,
\end{eqnarray}
where we used $p\cdot n=0$ and $\delta(p^2-m^2)p^2=\delta(p^2-m^2)m^2$, we find
\begin{eqnarray}
 \V^\mu&\!\!=&\!\!\frac{2}{(2\pi \hbar)^3}\sum_s\Bigg[\delta(p^2-m^2) \left(p^\mu-m s \frac\hbar2 \tilde{\omega}^{\mu\nu}
 n_\nu\frac{\partial}{\partial g_s^{(0)}}\right)\Bigg.\n\\ &&\left.+ ms \hbar \tilde{F}^{\mu\nu}n_\nu\delta'(p^2-m^2)\right.\n\\
 &&\Bigg.-\frac{s\hbar}{2m}\delta(p^2-m^2) \epsilon^{\mu\nu\alpha\beta}p_{\alpha} 
 \left(\nabla^{(0)}_\nu n_\beta\right)\Bigg]\n\\
 &&\times \left[\theta(p^0)f_s^{(0)+}+\theta(-p^0)f_s^{(0)-}\right]+\mathcal{O}(\hbar^2)\,. \label{V_eq}
\end{eqnarray}
The current given by Eq.\ (\ref{V_eq}) contains contributions which are not parallel to $p^\mu$. 
To first order in $\hbar$, particles are not transported parallel to their momenta.
The term containing $\tilde{F}^{\mu\nu}$ in Eqs.\ (\ref{V_eq}) is caused by off-shell effects and describes the
vector current induced by electromagnetic fields, which yields the analogue of the CME in the case of
non-zero mass. On the other hand, the term containing $\tilde{\omega}^{\mu\nu}$ describes 
the current induced by vorticity and thus gives the analogue of the CVE.

We furthermore calculate the axial-vector current. In order to do so, it is convenient to decompose the 
tensor $\bar{\Sigma}^{(1)}_{\mu\nu}$ introduced in Eq.\ \eqref{FA1} in the following way,
\begin{equation}
 \bar{\Sigma}^{(1)}_{\mu\nu}\equiv \frac12\chi_{\mu\nu} + \Xi_{\mu\nu}\,. \label{Sigmatilde}
\end{equation} 
The tensor $\Xi_{\mu\nu}$ is anti-symmetric and satisfies $p^\mu \Xi_{\mu\nu}=0$. On the other hand, 
$\chi_{\mu\nu}$ represents the dipole moment induced by the gradients of the distribution function since, 
according to Eq.\ \eqref{heyomega}, it satisfies
\begin{equation}
p^\nu \chi_{\mu\nu}=\nabla_\mu^{(0)}V^{(0)}\,. \label{omega}
\end{equation}   
Inserting $V^{(0)}$ into Eq.\ (\ref{omega}) and using Eq.\ \eqref{ooomega} we can derive the
following constraint for $\chi_{\mu\nu}$,
\begin{equation}
\label{cchhii}
p^\nu \chi_{\mu\nu}= p^\nu\omega_{\mu\nu} V^{(0)\prime} \, , 
\end{equation}
where we adopted the short-hand notation
\begin{equation}
V^{(0)\prime} \equiv \frac{2}{(2\pi \hbar)^3} 
\sum_s \frac{\partial}{\partial g_s^{(0)}} 
\left[\theta(p^0)f_s^{(0)+}+\theta(-p^0)f_s^{(0)-}\right] \, .
\end{equation}
The most general solution of Eq.\ \eqref{cchhii} can be written as
\begin{equation}\label{chi_def}
\chi_{\mu\nu}=\left[\kappa_1 \omega_{\mu\nu} - \frac{\kappa_2}{v \cdot p}(v_\mu\omega_{\nu\alpha}
-v_\nu\omega_{\mu\alpha})p^\alpha\right] V^{(0)\prime}\,,
\end{equation}
where $\kappa_{1,2}$ are arbitrary coefficients which satisfy $\kappa_1+\kappa_2=1$, and $v^\mu$ is an arbitrary 
vector such that $v \cdot p \neq 0$. Other possible terms which vanish when being contracted with the momentum are 
absorbed into $\Xi_{\mu\nu}$. The decomposition into $\chi_{\mu\nu}$ and $\Xi_{\mu\nu}$ is not unique, but allows for 
\textcolor{black}{the} transformations
\begin{eqnarray}
 \Xi_{\mu\nu}&\rightarrow& \Xi_{\mu\nu}-\frac C2[\omega_{\mu\nu}+\frac{1}{v \cdot p}(v_\mu\omega_{\nu\alpha}
 -v_\nu\omega_{\mu\alpha})p^\alpha]V^{(0)\prime} \, ,\n\\
 \chi_{\mu\nu} &\rightarrow& \chi_{\mu\nu}+C[\omega_{\mu\nu}+\frac{1}{v \cdot p}(v_\mu\omega_{\nu\alpha}
 -v_\nu\omega_{\mu\alpha})p^\alpha]V^{(0)\prime} \, ,\n\\
\end{eqnarray}
with $C$ being \textcolor{black}{an arbitrary} function of $x$ and $p$.
For any value of $\kappa_2$, we can apply the above transformation with $C\equiv\kappa_2$ to Eq.\ \eqref{chi_def}, 
which yields
 \begin{equation}
  \chi_{\mu\nu}=(\kappa_1+\kappa_2)\omega_{\mu\nu}V^{(0)\prime}=\omega_{\mu\nu}V^{(0)\prime}\,. \label{chi_phys}
 \end{equation}
Thus, we can set $\kappa_2=0$ without loss of generality. In other words, it is always possible to isolate the 
contribution proportional to $\omega_{\mu\nu}$ in the decomposition for $\bar{\Sigma}^{(1)}_{\mu\nu}$.  
This decomposition will \textcolor{black}{assume} 
a physical meaning when looking at the kinetic equation for $\Sc_{\mu\nu}$.

Inserting Eq.\ (\ref{Sigmatilde}) into Eq.\ (\ref{kin1}), we obtain
 \begin{eqnarray}
 0&=& p\cdot\nabla^{(0)} \left(\frac12\chi_{\mu\nu}+\Xi_{\mu\nu}\right)-F^\alpha_{\ [\mu}\left(\frac12\chi_{\nu]\alpha} 
 +\Xi_{\nu]\alpha}\right) \n\\
  &&  +\frac12(\partial_{x\alpha}F_{\mu\nu})\partial_p^{\alpha}V^{(0)} \, .
 \end{eqnarray}
Noting that $p\cdot\nabla^{(0)} \chi_{\mu\nu}=0$ and using Eqs.\ \eqref{relhommax} and \eqref{chi_phys}, 
we find that the $\chi_{\mu\nu}$-dependent part vanishes and
\begin{equation}
\label{mpdxi}
 p\cdot \nabla^{(0)} \Xi_{\mu\nu}=F^\alpha_{\  [\mu}\Xi_{\nu]\alpha}\, ,
\end{equation}
which is the second MPD equation for $\Xi_{\mu\nu}$. This part of the dipole-moment corresponds, 
together with the zeroth-order dipole moment, to the classical spin precession in electromagnetic fields. 

We now derive from Eq.~(\ref{PVA1}) the full axial-vector part of the Wigner function up to first order
\textcolor{black}{in $\hbar$}, i.e.,
 \begin{eqnarray}
\A^\mu&=&\frac{2}{(2\pi\hbar)^3}\sum_s\Bigg[\delta(p^2-m^2)  \Bigg.\n \\
 && \times \Bigg(s\,m\,n^{(0)\mu} -\frac{\hbar}{2} \tilde{\omega}^{\mu\nu}p_\nu
 \frac{\partial}{\partial g_s^{(0)}}\Bigg)
 \n\\ &&\Bigg.
 +\, \hbar \tilde{F}^{\mu\nu}p_\nu\delta'(p^2-m^2)\Bigg]\n\\
 &&\times \left[\theta(p^0)f_s^{(0)+}+ \theta(-p^0)f_s^{(0)-}\right] \n \\
&& -\frac\hbar2 \epsilon^{\mu\nu\alpha\beta}p_\nu \Xi_{\alpha\beta} \,\delta(p^2-m^2) +  \mathcal{O}(\hbar^2)\,. 
\label{A_eq2}
  \end{eqnarray}
By looking at the different terms in Eq.\ (\ref{A_eq2}), we identify three contributions to the axial-vector current in the 
massive case. The first term in the second line and the term in the last line describe the spin precession in 
\textcolor{black}{the} presence of \textcolor{black}{an} electromagnetic field according \textcolor{black}{to} 
the BMT equation. We remark that 
the function $\Xi_{\mu\nu}$ is not specified and has to be determined through Eq.\ \eqref{mpdxi}. The second term in the 
second line gives rise to the axial current in the direction of the vorticity, which is the analogue of the axial chiral vortical 
effect (ACVE). Finally, the term in the third line describes the axial current along the magnetic field, which is the analogue 
of the chiral separation effect (CSE). These terms are analogous to those found in Refs.\  
\cite{Fang:2016vpj,Becattini:2016gvu,Lin:2018aon}.

\section{Fluid-dynamical equations}

In this section, we present the equations of motion of the fluid-dynamical variables, i.e., of the
net particle-number current
and the energy-momentum tensor. We also give an equation for the spin tensor, which supplements these
equations in the case of spin-1/2 particles.

The net particle-number current is defined as
\begin{equation} \label{Jmu}
J^\mu(x)\equiv \langle:\bar{\psi}(x)\gamma^\mu \psi(x):\rangle=\int d^4 p \, \V^\mu(x,p)\,.
\end{equation}
Inserting the zeroth- and first-order
solutions (\ref{PVS0}), (\ref{PVA1}) into Eq.\ (\ref{Jmu}) we \textcolor{black}{obtain}
\begin{eqnarray}
J^\mu&=& \int dP \,p^\mu \left[V^{(0)}+\hbar V^{(1)}\right]+\frac{\hbar}{2}\partial_{x\nu}
\int dP \, \Sigma^{(0)\mu\nu}A^{(0)} \n\\
&&+\frac{\hbar}{4}F_{\alpha\beta}\int dP \, \partial_p^\mu \left[\Sigma^{(0)\alpha\beta}A^{(0)}\right]\,,
\end{eqnarray}
where $dP\equiv d^4 p\, \delta(p^2-m^2)$.

Equation (\ref{Vkin}) represents the conservation law for the vector component of the Wigner function.
Integrating this equation over kinetic 4-momentum, we immediately obtain the conservation law for the net particle-number
current,
\begin{equation}
\partial_{x\mu} J^\mu(x)=\int d^4 p \left[\nabla_{\mu} +j_0(\Delta)F_{\mu\nu}\partial_p^\nu\right]\V^\mu(x,p)=0\,,
\end{equation}
where we assumed that $F^{\mu\nu}$ is independent of $p^\nu$ and $\V^\mu$ vanishes sufficiently
rapidly for large momenta, which ensures the vanishing of a boundary term.

The Lagrangian operator for a Dirac spinor in an electromagnetic field is \cite{Vasak:1987um}
\begin{equation}
\mathcal{L}=\bar{\psi}\left[i \frac{\hbar}{2}\gamma \cdot \left(\overrightarrow{D}-\overleftarrow{D}^\dagger \right)
-m\right]\psi -\frac{1}{4}F^{\mu\nu}F_{\mu\nu}\,.
\end{equation}
From the Lagrangian we can derive the canonical energy-momentum tensor as follows,
\begin{eqnarray}
T^{\mu\nu}&=&\left\langle:\frac{\partial\mathcal{L}}{\partial(\partial_{x\mu} \psi)}\partial^\nu_x\psi
+\partial^\nu_x\bar{\psi}\frac{\partial\mathcal{L}}{\partial(\partial_{x\mu} \bar{\psi})}\right. \n\\
&&+ \left.\frac{\partial\mathcal{L}}{\partial(\partial_{x\mu} \mathbb{A}_\alpha)}\partial^\nu_x \mathbb{A}_\alpha
-g^{\mu\nu}\mathcal{L}:\right\rangle \n\\
&=&T^{\mu\nu}_{mat}+T^{\mu\nu}_{int} +T^{\mu\nu}_{em}\,,
\end{eqnarray}
where we have separated the total energy-momentum tensor into three parts: the gauge-invariant matter part
$T^{\mu\nu}_{mat}$, the part containing the interaction between gauge potential and matter current, $T^{\mu\nu}_{int}$,
and the electromagnetic part $T^{\mu\nu}_{em}$,
\begin{eqnarray}
T^{\mu\nu}_{mat}&=&\frac{\hbar}{2}\left\langle:\bar{\psi}\gamma^\mu (i\overrightarrow{D}^\nu
-i\overleftarrow{D}^{\dagger \nu})\psi :\right\rangle = \int d^4 p \, p^\nu \V^\mu\,, \n\\
T^{\mu\nu}_{int}&=& \mathbb{A}^\nu \left\langle:\bar{\psi} \gamma^\mu \psi:\right\rangle
=\mathbb{A}^\nu \int d^4 p\,  \V^\mu\,, \n\\
T^{\mu\nu}_{em}&=& \frac{1}{4}g^{\mu\nu}F^{\alpha\beta}F_{\alpha\beta}-F^{\mu\alpha}\partial^\nu_x \mathbb{A}_\alpha\,.
\label{Tmunu_s}
\end{eqnarray}
Note that none of these are in general symmetric under $\mu \leftrightarrow \nu$.
The total energy-momentum tensor is conserved $\partial_{x\nu} T^{\mu\nu}=0$, which can be checked using
the Dirac and Maxwell equations. However, the matter part is not conserved,
\begin{equation}
\partial_{x\mu} T^{\mu\nu}_{mat} (x)=F^{\nu\alpha}(x) J_\alpha(x)\,.
\end{equation}
This equation can be derived by acting $\partial_{x\mu}$ on $\V^\mu$ in the definition of $T^{\mu\nu}_{mat}$,
cf.\ first equation (\ref{Tmunu_s}), then using Eq.\ (\ref{Vkin}), and finally integrating by parts.
Inserting Eqs.\ (\ref{PVS0}), (\ref{PVA1}) into the energy-momentum tensor, we get
\begin{eqnarray}
T^{\mu\nu}_{mat}&=& \int dP\, p^\mu p^\nu\left[V^{(0)}+\hbar V^{(1)}\right]\n\\
&&+\frac{\hbar}{2}\partial_{x\alpha} \int dP\, p^\nu \Sigma^{(0)\mu\alpha}A^{(0)}\n\\
&& +\frac{\hbar}{4}g^{\mu\nu} F^{\alpha\beta}\int dP\, \Sigma_{\alpha\beta}^{(0)}A^{(0)}\n\\
&&-\frac{\hbar}{2}F^{\nu}_{\ \alpha}\int dP\, \Sigma^{(0)\mu\alpha}A^{(0)}  \n\\
&&+\frac{\hbar}{4}F^{\alpha\beta} \int dP \, p^\mu\partial_p^\nu \left[\Sigma_{\alpha\beta}^{(0)}A^{(0)}\right]\,.
\label{T_matter}
\end{eqnarray}

The total canonical angular momentum tensor is calculated as follows,
\begin{eqnarray}
J^{\lambda,\mu\nu}&=&x^\mu T^{\lambda\nu}-x^\nu T^{\lambda\mu}
+\frac{\hbar}{4}\left\langle:\bar{\psi}\{\gamma^\lambda,\sigma^{\mu\nu}\}\psi:\right\rangle \n\\
&& -(F^{\lambda\mu}\mathbb{A}^\nu-F^{\lambda\nu}\mathbb{A}^\mu)\,.
\end{eqnarray}
The first two terms, $x^\mu T^{\lambda\nu}-x^\nu T^{\lambda\mu}$, can be interpreted as the orbital
angular-momentum tensor. The remaining terms constitute the spin angular-momentum tensor, which can be further
separated into a matter and a field part. The spin tensor of matter can be defined as \cite{Becattini:2013fla}
\begin{eqnarray}
S_{mat}^{\lambda,\mu\nu}(x) & \equiv & \frac{1}{4}\left\langle:\bar{\psi}\{\gamma^\lambda, \sigma^{\mu\nu}\}\psi:
\right\rangle \n \\
&=& -\frac{1}{2}\epsilon^{\lambda\mu\nu\rho}\int d^4 p\, \A_\rho(x,p) \,.\label{S_A}
\end{eqnarray}
With the help of Eq.\ (\ref{Akin}) we find, to any order in $\hbar$,
\begin{equation}
\hbar\, \partial_{x\lambda} S^{\lambda,\mu\nu}_{mat}(x)=T^{\nu\mu}_{mat}(x)-T^{\mu\nu}_{mat}(x)\,, \label{S_matter}
\end{equation}
where we assumed that boundary terms vanish. Thus, the spin of matter is not conserved separately.
To zeroth order in $\hbar$, $T^{\mu\nu}_{mat}$ is symmetric according to Eq.\ (\ref{T_matter}), thus both sides of Eq.\
(\ref{S_matter}) vanish. To first order in $\hbar$, both sides are non-zero. Inserting the zeroth-order Wigner function
into Eq.\ (\ref{S_A}) we obtain
\begin{eqnarray}
S^{(0)\lambda,\mu\nu}_{mat}(x)& =& \frac{1}{2}\int dP \left[p^\lambda \Sigma^{(0)\mu\nu}
+ p^\mu \Sigma^{(0)\nu\lambda}\right. \n \\
&   & \hspace*{1.8cm} - \left. p^\nu \Sigma^{(0)\mu\lambda}\right] A^{(0)}\,. \label{Spin_tensor}
\end{eqnarray}
The above expressions for \textcolor{black}{the} energy-momentum and spin tensor emerge directly from 
Noether's theorem and thus correspond to the canonical ones. However, one can obtain different sets of 
tensors by applying pseudo-gauge transformations that keep the conservation laws for energy-momentum and spin. 
It has been shown that using different sets of tensors related through this pseudo-gauge freedom is not equivalent 
and leads to different measurable quantities \cite{Becattini:2018duy}.
We should mention that a similar derivation of fluid-dynamical equations of motion from the Wigner function for massless 
particles including the conservation of total angular momentum was carried out in Ref.~\cite{Yang:2018lew}.

Note that with Eq.\ (\ref{S_matter}), we can also prove that $\partial_{x\nu} T^{\mu \nu}_{mat} = F^{\mu \alpha} J_\alpha$,
the form of the equation of motion for the matter energy-momentum tensor given in 
Refs.\ \cite{Denicol:2018rbw,Denicol:2019iyh}.

\section{Conclusions}

In this paper we \textcolor{black}{have derived} kinetic theory for massive spin-1/2 particles in an 
inhomogeneous electromagnetic field
starting from the covariant formulation of the Wigner function. \textcolor{black}{Carrying out an expansion in $\hbar$ 
and truncating it at first order}, we found a general solution of the equations of motion. 
\textcolor{black}{We showed how to consistently take the massless limit and demonstrated agreement with 
previous works, which describe the CME and CVE}. One of the crucial
results of our work is the derivation of the collisionless Boltzmann equation for particles \textcolor{black}{that
carry a} dipole moment due to their
spin. We also recovered well-known results \textcolor{black}{in} the classical limit. The external force acting on the
particles is the sum of the Lorentz force and the Mathisson force, i.e., the first MPD equation.
The time evolution of the dipole
moment follows the second MPD equation, and the spin polarization precesses according to the BMT equation.
Moreover, as
an example, we studied the case of a rigidly rotating fluid in global equilibrium. In particular, we found the conditions that
the Lagrange multipliers related to the conservation of charge, energy, momentum, and angular momentum have to satisfy
in order for the distribution function to be a solution of the Boltzmann equation. Finally, fluid-dynamical equations
\textcolor{black}{of motion are provided,} in
which the spin tensor is included among the evolved densities.

\textcolor{black}{A straightforward extension of this work could be the inclusion of a collision term into our 
generalized Boltzmann equation and the derivation of the equations of motion for dissipative relativistic
magneto-hydrodynamics for spin-1/2 particles. This could be achieved using the method of moments, following
Refs.\ \cite{Denicol:2018rbw,Denicol:2019iyh}, where this has already been done for spin-0 particles.
Another potential extension would be the derivation of a transport equation starting from the
equal-time Wigner-function formalism \cite{Zhuang:1995pd}.}

\section*{Acknowledgements}

The authors thank F.\ Becattini, W.\ Florkowski, C.\ Greiner, K.\ Hattori, U.\ Heinz, X.-G.\ Huang, E.\ Moln\'{a}r,
L.\ Tinti, and H.\ van Hees for enlightening discussions.
The work of D.H.R., X.-l.S., E.S., and N.W.\ is supported by the
Deutsche Forschungsgemeinschaft (DFG, German Research Foundation)
through the Collaborative Research Center CRC-TR 211 ``Strong-interaction matter
under extreme conditions'' -- project number 315477589 - TRR 211. D.H.R.\ acknowledges partial support
by the High-end Foreign Experts project GDW20167100136 of the State
Administration of Foreign Experts Affairs of China.
X.-l.S.\ is supported in part by China Scholarship Council. E.S.\ acknowledges support by BMBF
``Verbundprojekt: 05P2015 - ALICE at High Rate", and 
BMBF ``Forschungsprojekt: 05P2018 - Ausbau von ALICE am LHC (05P18RFCA1)".
Q.W.\ is supported in part by the 973 program under
Grant No.\ 2015CB856902 and by NSFC under Grant No.\ 11535012.

\section*{Note added}
After completion of this work, we became aware of a related study \cite{Gao:2019znl}, where kinetic equations for
massive fermions were derived using the covariant Wigner-function approach. Other related work, which
appeared after the submission of this paper, can be found in Refs.~\cite{Hattori:2019ahi,Wang:2019moi}.

\appendix
\section{Diagonal spin basis}\label{XX}

In this appendix, we show how to diagonalize the distribution function by choosing the
spin quantization direction along the polarization direction.
The axial-vector current which one obtains directly from Eq.\ (\ref{solution_hbar=0}) reads
 \begin{eqnarray}
\lefteqn{  \A^{(0)}_\mu (x,p) = \frac{1}{(2\pi \hbar)^3}\delta(p^2-m^2) }\n \\
  & \times & \sum_{rs}
  \left[\theta(p^0)\bar{u}(\mathbf{p},s)\gamma_\mu\gamma^5 u(\mathbf{p},r)f^{(0)+}_{rs}(x,\mathbf{p})
  \right. \n \\
  &   - & \left. \theta(-p^0)\bar{v}(-\mathbf{p},r)\gamma_\mu\gamma^5 v(-\mathbf{p},s)f_{sr}^{(0)-}(x,-\mathbf{p})
  \right]. \label{A_nondiag}
 \end{eqnarray}
The
distribution functions $f^{(0)\pm}$ are Hermitian matrices in spin space and can thus be diagonalized by a unitary
transformation \cite{DeGroot:1980dk}. Since the Pauli matrices $\sigma^i$ together with the unit matrix
are a basis of the space of Hermitian $(2 \times2)$ matrices, the distribution functions can be written as \cite{Leader:2001,Florkowski:2017dyn}
\begin{equation}
 f^{(0)e} = a^e+\mathbf{b}^e \cdot \textrm{\boldmath$\mathbf{\sigma}$}\,,
\end{equation}
with some coefficients $a^e$ and $\mathbf{b}^e$ and $e=\pm$ represents positive-/negative-energy states.

In the rest frame, the standard spinors $u$ and $v$ are given as \cite{itzykson2012quantum}
\begin{eqnarray}
 u(\mathbf{0},+)&=&\sqrt{2m}\; (1,0,0,0)^T \,, \n\\
 u(\mathbf{0},-)&=&\sqrt{2m}\; (0,1,0,0)^T \,, \n\\
 v(\mathbf{0},+)&=&\sqrt{2m}\; (0,0,1,0)^T \,, \n\\
 v(\mathbf{0},-)&=&\sqrt{2m}\; (0,0,0,1)^T\,. \label{uvrf}
\end{eqnarray}
Note that $u(\mathbf{0},+)$ corresponds to a particle with spin parallel to the $z$-direction, while $v(\mathbf{0},+)$
corresponds to an anti-particle with spin anti-parallel to the $z$-direction.

We diagonalize the distribution functions $f^{(0)e}$ in the rest frame,
\begin{equation}
 {f}^{(0)e}_{s}\delta_{rs}=\sum_{r^\prime s^\prime}(D^e)^\dagger_{rr^\prime} f^{(0)e}_{r's'} D^e_{s^\prime s}\,, \label{diag_f}
\end{equation}
with $D^e$ being $2\times 2$ matrices in spin space,
\begin{equation}
 D^e=\left(\vec{d}^e_+,\vec{d}^e_-\right)\,,
\end{equation}
where $\vec{d}^e_\pm$ are the eigenvectors of $\mathbf{b}^e\cdot \textrm{\boldmath$\mathbf{\sigma}$}$
corresponding to the eigenvalues $\pm$, respectively,
\begin{equation} \label{spinqu}
\left(\mathbf{n}^e\cdot \textrm{\boldmath$\mathbf{\sigma}$}\right) \vec{d}^e_\pm = \pm e \, \vec{d}^e_\pm\,,
\end{equation}
where $\mathbf{n}^e\equiv \mathbf{b}^e/\sqrt{\mathbf{b}^e\cdot\mathbf{b}^e}$ is the unit vector along the
direction of $\mathbf{b}^e$.
Note that the distribution functions $f^{(0)e}$ in general depend on the space-time coordinates ${x^\mu}$,
thus the transformation matrices $D^e$ as well as $\mathbf{n}^e$ are defined locally. We then define the following
spinors, which can be derived by rotating the standard ones,
\begin{eqnarray}
&&\tilde{u}(x,\mathbf{0},s)\equiv \sum_{s^\prime} u(\mathbf{0},s^\prime) D^+_{s^\prime s}(x) = \sqrt{2m}\left(\begin{array}{c}
           \vec{d}^+_s \\
           \vec{0}
         \end{array}\right)\,, \n\\
&&\tilde{v}(x,\mathbf{0},s)\equiv \sum_{s^\prime} v(\mathbf{0},s^\prime) D^-_{s^\prime s}(x) = \sqrt{2m}\left(\begin{array}{c}
           \vec{0} \\
           \vec{d}^-_s
         \end{array}\right)\,. \n\\ \label{tilde_u}
\end{eqnarray}
The spinors $\tilde{u}(x,\mathbf{0},\pm)$ and $\tilde{v}(x,\mathbf{0},\pm)$ now correspond to particles/anti-particles
with spin parallel/anti-parallel to $\mathbf{n}^\pm$.
Using Eqs.\ (\ref{diag_f}) and (\ref{tilde_u}) we obtain
\begin{eqnarray}
\lefteqn{\sum_{rs} \bar{u}(\mathbf{0},s) \gamma^\mu\gamma^5 u(\mathbf{0},r) f^{(0)+}_{rs}(x,\mathbf{0})}\n\\
&=& \sum_{rs} \bar{\tilde{u}}(x,\mathbf{0},s) \gamma^\mu\gamma^5 \tilde{u}(x,\mathbf{0},r)
f^{(0)+}_{s}(x,\mathbf{0}) \delta_{rs}\n\\
&=& 2m \sum_s s \left(0, \mathbf{n}^+\right)f^{(0)+}_{s}(x,\mathbf{0}) \,,
\end{eqnarray}
and similarly for the $v$--spinors.
Then, performing a Lorentz transformation we obtain
\begin{eqnarray}
\lefteqn{\sum_{rs} \bar{u}(\mathbf{p},s) \gamma^\mu\gamma^5 u(\mathbf{p},r) f^{(0)+}_{rs}(x,\mathbf{p})}\n\\
&=& 2m \sum_s s\, n^{+\mu}(x,\mathbf{p},\mathbf{n}^+) f^{(0)+}_{s}(x,\mathbf{p})\,, \label{u_identity}
\end{eqnarray}
and similarly
\begin{eqnarray}
\lefteqn{\sum_{rs} \bar{v}(-\mathbf{p},s) \gamma^\mu\gamma^5 v(-\mathbf{p},r) f^{(0)-}_{rs}(x,-\mathbf{p})}\n\\
&=& 2m \sum_s s\, n^{-\mu}(x,-\mathbf{p},-\mathbf{n}^-) f^{(0)-}_{s}(x,-\mathbf{p})\,, \label{v_identity}
\end{eqnarray}
where $n^{\pm\mu}$ is given by Eq.\ (\ref{nDef}).
We rewrite the axial-vector current as
\begin{equation}
 \A^{(0)}_\mu=m n^{(0)}_\mu A^{(0)}\delta(p^2-m^2)\,, \label{A_final}
\end{equation}
where the vector $n^{(0)}_\mu(x,p)$ and the distribution function $A^{(0)}(x,p)$ are determined by Eqs.\
(\ref{nmu}) and (\ref{A_Wignerdirect}), respectively.

Furthermore, we define
\begin{equation}
\label{sigg1}
 s\Sigma^{+ \mu\nu}(x, \mathbf{p})\equiv \frac{1}{2m}\bar{\tilde{u}}(x,\mathbf{p},s)\sigma^{\mu\nu}\tilde{u}(x,\mathbf{p},s)\,,
\end{equation}
and
\begin{equation}
\label{sigg2}
 s\Sigma^{-\mu\nu}(x,\mathbf{p})\equiv \frac{1}{2m}\bar{\tilde{v}}(x,-\mathbf{p},s)\sigma^{\mu\nu}
 \tilde{v}(x,-\mathbf{p},s)\,.
\end{equation}
We have
\begin{equation}
 \Sigma^{\pm\mu\nu}(x,\mathbf{p})=-\frac{1}{m} \epsilon^{\mu\nu\alpha\beta}p_\alpha
 n^\pm_\beta(x,\mathbf{p})\, ,
\end{equation}
which can be easily checked in the rest frame using the Dirac representation of the
$\gamma$--matrices and $[\sigma_i,\sigma_j]=2i\epsilon_{ijk}\sigma_k$. Defining
\begin{equation}
 \Sigma^{(0)\mu\nu}(x, p)\equiv \theta(p^0)\Sigma^{+\mu\nu}(x,\mathbf{p})
 -\theta(-p^0)\Sigma^{-\mu\nu}(x,\mathbf{p})\,, \label{Sigma0}
\end{equation}
we obtain the tensor current $\Sc^{(0)}_{\mu\nu}$ as
\begin{eqnarray}
\lefteqn{ \Sc_{\mu\nu}^{(0)}(x,p) =\frac{1}{(2\pi \hbar)^3}\delta(p^2-m^2) }\n \\
  & \times & \sum_{s}
 \left[\theta(p^0)\bar{\tilde{u}}(x,\mathbf{p},s)\sigma_{\mu\nu} \tilde{u}(x,\mathbf{p},s){f}^{(0)+}_{s}(x,\mathbf{p}) \right. \n \\
 & & -\left. \theta(-p^0)\bar{\tilde{v}}(-\mathbf{p},s)\sigma_{\mu\nu} \tilde{v}(-\mathbf{p},s){f}_{s}^{(0)-}(x,-\mathbf{p})\right]\n\\
 &=& m\,\Sigma^{(0)}_{\mu\nu}(x, p)\delta(p^2-m^2)A^{(0)}(x,p)\,.
\end{eqnarray}
Using
\begin{eqnarray}
 \bar{\tilde{u}}(x,\mathbf{p},s)\tilde{u}(x,\mathbf{p},s)&=&-\bar{\tilde{v}}(x,-\mathbf{p},s)\tilde{v}(x,-\mathbf{p},s)
 \n \\ & = & 2m\,,\n\\
 \bar{\tilde{u}}(x,\mathbf{p},s)\gamma^5\tilde{u}(x,\mathbf{p},s)
 &=&\bar{\tilde{v}}(x,-\mathbf{p},s)\gamma^5\tilde{v}(x,-\mathbf{p},s)\n \\
 &=& 0\,,\n\\
 \bar{\tilde{u}}(x,\mathbf{p},s)\gamma^\mu\tilde{u}(x,\mathbf{p},s)
 &=&-\bar{\tilde{v}}(x,-\mathbf{p},s)\gamma^\mu\tilde{v}(x,-\mathbf{p},s)\n \\
 &= &2p^\mu \,,
\end{eqnarray}
the calculation of $\F^{(0)}$, $\Pc^{(0)}$, and $\V^{(0)}_\mu$ is straightforward. 

Finally, we  stress that the diagonalization procedure for the distribution function described in this appendix 
is in general possible also at higher order in $\hbar$, even though the exact form of the spinors is not known.

\section{Redundancy of Eqs.\ (\ref{F}) -- (\ref{Akin})}\label{prove}

In this section we prove that Eqs.\ (\ref{F}) -- (\ref{Akin}) are not independent from each other.
Combining Eqs.\ (\ref{F}), (\ref{S}), (\ref{Vkin}), and (\ref{Akin}), we derive
\begin{eqnarray}
0&=&\frac{\hbar}{2m}\nabla_\mu\left(\Pi \cdot \V-m\F \right)-\frac{1}{2m}\Pi_\mu\left(\hbar\nabla \cdot \V \right)\n\\
&&-\frac{1}{m}\Pi^\nu\left(\frac{\hbar}{2}\nabla_{[\mu} \V_{\nu]}-\epsilon_{\mu\nu\alpha\beta}\Pi^\alpha \A^\beta
-m\Sc_{\mu\nu}\right)\n\\
&&+\frac{\hbar}{2m}\nabla^\nu\left(\Pi_{[\mu}\V_{\nu]}+\frac{\hbar}{2}\epsilon_{\mu\nu\alpha\beta}
\nabla^\alpha \A^\beta\right)\,.
\end{eqnarray}
After some calculation we obtain
\begin{eqnarray}
&&\frac{\hbar}{2}\nabla_\mu\F+\Pi^\nu\Sc_{\nu\mu} \n\\
&=&\frac{\hbar}{2m}\left([\nabla_\mu,\Pi_\nu]+[\nabla_\nu,\Pi_\mu]\right)\V^\nu
+\frac{\hbar}{2m}[\Pi^\nu,\nabla_\nu]\V_\mu \n\\
&&+\frac{1}{2m}\epsilon_{\mu\nu\alpha\beta}\left([\Pi^\nu,\Pi^\alpha]
+\frac{\hbar^2}{4}[\nabla^\nu,\nabla^\alpha]\right)\A^\beta\,.\label{nablaF}
\end{eqnarray}
The commutators can be easily calculated using the definition of the operators (\ref{pW}):
\begin{eqnarray}
\left[\Pi_\mu,\Pi_\nu\right] &=& -\hbar j_{1}(\Delta)F_{\mu\nu}-\frac{\hbar}{2}\Delta j_{1}^{\prime}(\Delta)F_{\mu\nu}\,,\n\\
\left[\Pi_\mu,\nabla_\nu\right] &=& \Delta j_1(\Delta)F_{\mu\nu}-j_0(\Delta)F_{\mu\nu}\,,\n\\
\left[\nabla_\mu,\nabla_\nu\right] &=& \frac{2}{\hbar}\Delta j_0(\Delta)F_{\mu\nu}\,,
\end{eqnarray}
where $j_1^{\prime}(x)\equiv\frac{d}{dx}j_1(x)$. Using the definitions of the spherical Bessel functions we can prove
\begin{equation}
xj_0(x)-2j_1(x)-xj_1^{\prime}(x)=0\,.
\end{equation}
Inserting the commutators into Eq.\ (\ref{nablaF}) and using the above relation, one finds that the right-hand
side of Eq.\ (\ref{nablaF}) vanishes, and we just obtain Eq.\ (\ref{B}).

Analogously, we can construct the following equation using Eqs.\ (\ref{P}), (\ref{S}), (\ref{orth}), and (\ref{Akin}),
\begin{eqnarray}
0&=&\frac{1}{m}\Pi_\mu\left(\frac{\hbar}{2}\nabla \cdot \A +m\Pc\right)-\frac{\hbar}{2m}\nabla_\mu\left( \Pi \cdot \A \right) \n\\
&&-\frac{\hbar}{4m}\epsilon_{\mu\nu\alpha\beta} \nabla^\nu \n\\
&&\quad\times\left[\frac{\hbar}{2}\left(\nabla^\alpha\V^\beta-\nabla^\beta\V^\alpha\right)
-\epsilon^{\alpha\beta\rho\sigma}\Pi_\rho \A_\sigma-m\Sc^{\alpha\beta}\right]\n\\
&&-\frac{1}{2m}\epsilon_{\mu\nu\alpha\beta}\Pi^\nu\left(\Pi^\alpha\V^\beta-\Pi^\beta\V^\alpha
+\frac{\hbar}{2}\epsilon^{\alpha\beta\rho\sigma}\nabla_\rho\A_\sigma\right)\,,\n\\
\end{eqnarray}
from which we get
\begin{eqnarray}
&& \Pi_\mu \Pc+\frac{\hbar}{4}\epsilon_{\mu\nu\alpha\beta}\nabla^\nu\Sc^{\alpha\beta}\n\\
&=& -\frac{\hbar}{2m}\left([\Pi_\mu,\nabla_\nu]+[\Pi_\nu,\nabla_\mu]\right)\A^\nu
+\frac{\hbar}{2m}[\Pi^\nu,\nabla_\nu]\A_\mu \n\\
&&+\frac{1}{2m}\epsilon_{\mu\nu\alpha\beta}\left([\Pi^\nu,\Pi^\alpha]
+\frac{\hbar^2}{4}[\nabla^\nu,\nabla^\alpha]\right)\V^\beta\,. \label{PiP}
\end{eqnarray}
Analogously to Eq.\ (\ref{nablaF}), the right-hand side of Eq.\ (\ref{PiP}) vanishes and we obtain Eq.\ (\ref{Skin}).

\section{Derivation of kinetic equations}\label{appendixB}

In this appendix we show some technical details we used when deriving the kinetic equations (\ref{kin0}), (\ref{kin1}).
First we focus on the kinetic equation for the zeroth-order dipole-moment tensor $\Sigma^{(0)}_{\mu\nu}$ and
the axial distribution function $A^{(0)}$. We insert the vector part of Eq.\ (\ref{PVA1_org}) into Eq.\ (\ref{constrain1})
and use the relation $\A_\mu^{(0)} = -\frac{1}{2m}\epsilon_{\mu\nu\alpha\beta}p^\nu \Sc^{\alpha\beta(0)}$, to derive
\begin{eqnarray}
0&=&-\frac{1}{2m}\left[p_\mu \nabla^{(0)\alpha}\Sc_{\alpha\nu}^{(0)}-p_\nu \nabla^{(0)\alpha}\Sc_{\alpha\mu}^{(0)}\right] \n\\
&&-\frac{1}{4m}\epsilon_{\mu\nu\alpha\beta}\nabla^{(0)\alpha} \epsilon^{\beta\gamma\rho\sigma}p_\gamma
\Sc_{\rho\sigma}^{(0)} \n\\
&=& -\frac{1}{2m}\left\{ \left[\nabla^{(0)\alpha}p_{[\mu}\right]\Sc_{\nu]\alpha}^{(0)}
+p^\alpha\nabla_{\alpha}^{(0)} \Sc^{(0)}_{\mu\nu}\right\}\,.
\end{eqnarray}
Inserting the zeroth-order solution we get
\begin{equation}
\delta(p^2-m^2)\left\{p \cdot \nabla^{(0)}\left[\Sigma^{(0)}_{\mu\nu}A^{(0)} \right]
-F_{\ [\mu}^\alpha\Sigma^{(0)}_{\nu]\alpha}A^{(0)}\right\}=0\,. \label{eqSigmaA}
\end{equation}
The dipole-moment tensor is normalized, $\Sigma^{(0)\mu\nu}\Sigma^{(0)}_{\mu\nu}=2$,
thus contracting the above equation with $\Sigma^{(0)\mu\nu}$ we obtain
\begin{equation}
\delta(p^2-m^2)p \cdot \nabla^{(0)}A^{(0)}=0\,, \label{eqA}
\end{equation}
where we have used
\begin{equation}
\Sigma^{(0)\mu\nu}\left[F_{\ \mu}^\alpha\Sigma^{(0)}_{\nu\alpha}-F_{\ \nu}^\alpha\Sigma^{(0)}_{\mu\alpha}\right]
=2F_{\alpha\mu}\Sigma^{(0)\mu\nu}\Sigma^{(0)\alpha}_{\nu}=0\,,
\end{equation}
because $F_{\alpha\mu}$ is anti-symmetric and $\Sigma^{(0)\mu\nu}
\Sigma^{(0)\ \alpha}_{\nu}$ is symmetric under $\alpha \leftrightarrow \mu$.
Inserting Eq.\ (\ref{eqA}) into Eq.\ (\ref{eqSigmaA}) one obtains the kinetic equation for $\Sigma^{(0)}_{\mu\nu}$.

The kinetic equation for $V^{(1)}$ is derived from the first line of Eq.\ (\ref{constrain2}).
According to Eq.\ (\ref{PVA1_org}), $\V^{(1)}$ can be expressed in terms of $\F^{(1)}$ and $\Sc_{\mu\nu}^{(0)}$.
Thus we get
\begin{equation}
\frac{1}{m}p \cdot \nabla^{(0)}\F^{(1)}+\frac{1}{2m}\nabla^{(0)\mu}\nabla^{(0)\nu}\Sc_{\mu\nu}^{(0)}=0\,.
\end{equation}
The dipole-moment tensor is anti-symmetric in its indices, so we can use the commutator
$[\nabla^{(0)\mu},\nabla^{(0)\nu}]=(\partial^\alpha_x F^{\mu\nu})\partial_{p\alpha}$ to simplify the second term.
Using also the zeroth- and first-order solutions we obtain
\begin{eqnarray}
0&=&\frac{1}{m}p\cdot \nabla^{(0)}\F^{(1)}+\frac{1}{4m}(\partial_x^\alpha F^{\mu\nu})\partial_{p\alpha}
\Sc_{\mu\nu}^{(0)} \n\\
&=& \delta(p^2-m^2)p \cdot \nabla^{(0)}V^{(1)} \n\\
&&-\frac{1}{2}\delta^\prime(p^2-m^2)p \cdot \nabla^{(0)}\left[F^{\alpha\beta}\Sigma_{\alpha\beta}^{(0)}A^{(0)}
\right] \n\\
&& +\frac{1}{4}(\partial_x^\alpha F^{\mu\nu})\partial_{p\alpha} \left[\Sigma_{\mu\nu}^{(0)}A^{(0)}\delta(p^2-m^2)
\right] \n\\
&=& \delta(p^2-m^2)\left\{p \cdot \nabla^{(0)}V^{(1)} \right. \n \\
& & \hspace*{1.3cm} +\left. \frac{1}{4}(\partial_x^\alpha F^{\mu\nu})
\partial_{p\alpha} \left[\Sigma_{\mu\nu}^{(0)}A^{(0)}\right]\right\} \n\\
&&-\frac{1}{2}\delta^\prime(p^2-m^2)F^{\alpha\beta}p \cdot \nabla^{(0)}\left[\Sigma_{\alpha\beta}^{(0)}A^{(0)}\right]\,.
\end{eqnarray}

In order to derive the kinetic equation for the first-order dipole-moment tensor, we first need
$\V^{(2)}_\mu$, which is calculated by expanding Eq.\ (\ref{PVS}) into a series in $\hbar$ and
identifying the $\hbar^2$ term,
\begin{equation}
\V^{(2)}_\mu=\frac{1}{m}p_\mu\F^{(2)}+\frac{1}{m}\Pi^{(2)}_\mu\F^{(0)}+\frac{1}{2m}\nabla^{(0)\nu}\Sc_{\mu\nu}^{(1)}\,.
\end{equation}
Inserting this, as well as $\A_\mu^{(1)}$ from Eq.\ (\ref{PVA1_org})
into the second line of Eq.\ (\ref{constrain2}) we get
\begin{eqnarray}
0&=&\frac{1}{m}\left([\Pi_\mu^{(2)},p_\nu]-[\Pi_\nu^{(2)},p_\mu]\right)\F^{(0)}
-\frac{1}{2m} p\cdot \nabla^{(0)}\Sc^{(1)}_{\mu\nu}\n\\
&& +\frac{1}{2m}\left([\nabla^{(0)\alpha},p_\mu]\Sc_{\alpha\nu}^{(1)}
-[\nabla^{(0)\alpha},p_\nu]\Sc_{\alpha\mu}^{(1)}\right)\,. \label{kin_Sigma1}
\end{eqnarray}
The commutators are given by $[\nabla^{(0)\alpha},p_\mu]=F_\mu^{\ \alpha}$ and
\begin{equation}
[\Pi_\mu^{(2)},p_\nu]-[\Pi_\nu^{(2)},p_\mu]= -\frac{1}{4}(\partial_{x\alpha} F_{\mu\nu})\partial_p^\alpha\,.
\end{equation}
Inserting the solutions for $\F^{(0)}$ and $\Sc_{\mu\nu}^{(1)}$
from Eqs.\  (\ref{FA0}) and (\ref{FA1}) into Eq.\ (\ref{kin_Sigma1}) and using the above commutators, one obtains
the kinetic equation for $\bar{\Sigma}_{\mu\nu}^{(1)}$.

\bibliography{biblio_paper}{}

\end{document}